# Many-body Expansion Based Machine Learning Models for Octahedral Transition Metal Complexes

Ralf Meyer[1], Daniel B. K. Chu[1], and Heather J. Kulik[1,2,*]

[1]*Department of Chemical Engineering, Massachusetts Institute of Technology, Cambridge, MA 02139, USA*

[2]*Department of Chemistry, Massachusetts Institute of Technology, Cambridge, MA 02139, USA*

ABSTRACT:

Graph-based machine learning models for materials properties show great potential to accelerate virtual high-throughput screening of large chemical spaces. However, in their simplest forms, graph-based models do not include any 3D information and are unable to distinguish stereoisomers such as those arising from different orderings of ligands around a metal center in coordination complexes. In this work we present a modification to revised autocorrelation descriptors, our molecular graph featurization method for machine learning various spin state dependent properties of octahedral transition metal complexes (TMCs). Inspired by analytical semi-empirical models for TMCs, the new modeling strategy is based on the many-body expansion (MBE) and allows one to tune the captured stereoisomer information by changing the truncation order of the MBE. We present the necessary modifications to include this approach in two commonly used machine learning methods, kernel ridge regression and feed-forward neural networks. On a test set composed of all possible isomers of binary transition metal complexes, the best MBE models achieve mean absolute errors of 2.75 kcal/mol on spin-splitting energies and 0.26 eV on frontier orbital energy gaps, a 30-40% reduction in error compared to models based on our previous approach. We also observe improved generalization to previously unseen ligands where the best-performing models exhibit mean absolute errors of 4.00 kcal/mol (i.e., a 0.73 kcal/mol reduction) on the spin-splitting energies and 0.53 eV (i.e., a 0.10 eV reduction) on the frontier orbital energy gaps. Because the new approach incorporates insights from electronic structure theory, such as ligand additivity relationships, these models exhibit systematic generalization from homoleptic to heteroleptic complexes, allowing for efficient screening of TMC search spaces.

# 1. Introduction.

Virtual high-throughput screening (VHTS) has emerged as a valuable tool for computational discovery in many fields of material science, such as the design of catalysts,[1-5] metal–organic frameworks,[6-10] and components for energy storage.[11-15] One particularly interesting class of materials for VHTS are transition metal complexes (TMCs) due to their highly tunable properties and large chemical design spaces with potential applications ranging from optoelectronic devices[16-18] to biomedicine.[19-22] The computational method of choice for these studies has long been density functional theory (DFT) due to its accuracy at reasonable cost. However, as researchers are tackling ever larger chemical design spaces, machine learning surrogate models trained on DFT or higher-accuracy data sets have started to establish themselves as a lower-cost alternative to DFT for VHTS.[23-27] Depending on the application and its requirements regarding accuracy and computational cost, these ML models can be either drop-in replacements for DFT, i.e., mappings from the atomic structure to material properties,[28-32] or operate on an abstracted representation of the design space, such as the molecular graph[33-37] or the crystal graph.[38-42] The latter approach allows one to perform VHTS in the abstracted representation space, bypassing the costly steps of building and optimizing a 3D geometry, and thereby enabling the screening of millions of candidates in a matter of minutes. Nevertheless, even for a moderately sized set of potential ligands, the combinatorial explosion of possible arrangements into coordination complexes gives rise to a vast design space.[43-47] On the other hand, automated DFT evaluation of open-shell transition metal systems is difficult due to potential convergence issues and the presence of multi-reference character, severely limiting the size of available data sets.

The large number of theoretical complexes and relative paucity of studied complexes necessitates a trade-off between expressivity and data efficiency in the design of TMC descriptors for machine learning. That is, to enable model training on small data sets, the descriptors compress the structural information as much as possible. The process of encoding atomic structures using higher-level representations and featurization of those representations to obtain fixed-size descriptors for ML models necessarily comes with a loss of information. However, this loss of information limits the potential accuracy of the ML models. For the specific use case of TMCs, describing them only by their molecular graph in machine learning also gives rise to challenges. The central metal ion in a TMC allows for multiple coordination geometries, a fact that is not captured by the molecular graph representation. Even when the design space is limited to a single coordination geometry, certain combinations of ligands allow for multiple stereoisomers, such as the *cis/trans* and *fac/mer* isomers of binary octahedral TMCs. For many applications, this stereoisomerism results in significantly different TMC properties. Notable examples include the anticancer drug cisplatin for which the *trans* isomer is clinically ineffective[48-51] or stereoisomer-dependent reactivities of TMC catalysts.[52-54] It is, therefore, crucial to encode this minimal amount of 3D information in the featurization of the molecular graph.

We have previously addressed these challenges with revised autocorrelations (RACs),[55] a set of molecular graph descriptors tailored to mononuclear octahedral TMCs. However, in its original implementation, this approach requires an additional step of assigning an equatorial plane of ligands and results in discontinuities in the feature space for certain ligand compositions and an inability to distinguish stereoisomers in some edge cases. In analytical models for TMC properties such as ligand field parameterization schemes,[56,57] Bursten additivity,[58] and ligand field molecular mechanics,[59] which were derived from early semi-empirical electronic structure theory approaches,[60,61] these stereoisomer effects are modeled by dividing the metal–ligand interaction into nonadditive and additive ligand field contributions. A unifying mathematical framework for these methods is given by the many-body expansion (MBE). Inspired by the success of the MBE in molecular modeling[62-67] and, more recently, in machine learning interatomic potentials,[68-72] we present a MBE-based approach for featurizing the molecular graph of octahedral TMCs. Combined with appropriate modifications to the ML models, this method allows one to systematically address the challenges of encoding octahedral stereoisomerism and the accuracy/data-efficiency trade-off by varying the truncation order of the MBE. We derive the necessary expressions to include this approach into two commonly used ML models, kernel ridge regression (KRR) and neural networks (NN), and compare the performance of these new models to our previous approach using spin-splitting energies and frontier orbital energies as target properties.

**2. Many-body expansion based machine learning models.**

**2.1 Many-body expansion of metal-centric properties.**

The influence of ligands on a metal-centric property $Q$ of an octahedral transition metal complex (TMC) can be expanded in a series of $n$-body interaction terms, $q_n$:

$$Q = q_1(M) + \sum_{i=1}^{6} q_2(M, L_i) + \sum_{i=1}^{6}\sum_{j>i}^{6} q_3(M, L_i, L_j) + \cdots, \quad (1)$$

where $M$ encodes the identity, oxidation, and spin state of the central metal ion and $L_i$ refers to the identity of the ligand $i$. Following recent examples[71-73], we investigate the use of machine learning (ML) to model the individual interaction terms in the many-body expansion (MBE). Here, we limit our analysis to two model types corresponding to truncation of the MBE at the two-body and three-body interaction level.

Truncation at the two-body interaction terms give rise to the concept of ligand additivity, which has been shown to be useful for the description of many properties of TMCs.[47,56,74,75] However, in the simplest models that are restricted to single-ligand–metal interactions, there is an inability to distinguish different structural isomers for the same ligand composition (e.g., *cis* versus *trans* orientations), thereby limiting the applicability of such models. A notable example of models of this complexity is the DBLOC scheme for correcting DFT (e.g., B3LYP) spin-splitting energy predictions with respect to experimental reference values.[76] In DBLOC, the strength of the two-body interaction term depends on the identity of the metal, the oxidation state, and the coordinating atom type of each ligand.

Inclusion of the three-body order yields 15 additional interaction terms for the case of a mononuclear octahedral TMC. Depending on the relative positioning of the two ligands involved, these interactions can be divided into 12 *cis*-type and 3 *trans*-type interactions. Neglecting the geometry dependence of the interaction terms beyond *cis* and *trans* positioning leads to linear dependence between the set of two-body and three-body interactions (Appendix A). In this work, we make no modification to our ML models to account for this linear dependence. Note, however, that for certain ML methods it might be necessary to resolve the linear dependence prior to fitting the model.

**2.2 Featurization**

To simplify comparisons to our previous work, we base the featurization for the machine-learned interaction terms on the revised autocorrelations (RACs) featurization for TMCs.[55] RACs expand on the idea of autocorrelations (ACs)[77] of atomic properties on the molecular graph by calculating correlations between restricted subsets of atoms:

$$_{\text{scope}}^{\text{start}}P_d = \sum_{i \in \text{start}} \sum_{j \in \text{scope}} P_i P_j \delta(d_{ij}, d), \tag{1}$$

where start and scope refer to the two involved subsets of atoms, $d$ is the depth of the descriptor, $P_i$ is an atomic property of atom $i$, $\delta$ is the Kronecker delta, and $d_{ij}$ is the path distance between atoms *i* and *j* on the molecular graph. The five most commonly computed heuristic properties are nuclear charge, Z, Pauling electronegativity, χ, topology, T, which is the atom's coordination number, identity, I, which is 1 for any atom, and covalent atomic radius, S. In this work these correlation functions are evaluated for four different combinations of start/scope atom sets and up to a depth of three bonds, i.e., $d=3$. The first two start/scope combinations correspond to standard ACs where both start and scope include all atoms (all). One may also compute metal-centered (mc) descriptors where the start set is restricted to just metal atoms while the scope still includes all atoms. Both of these sets of descriptors apply to the whole TMC. For the other two combinations, the scope set is always limited to all atoms of a single ligand (lig) and the start subset includes either all atoms of the ligand or just the connecting atom (lc). In octahedral systems the descriptors of only the individual ligands $_{\text{lig}}^{\text{lig}}P_d$ and $_{\text{lig}}^{\text{lc}}P_d$ are separately averaged over the four equatorial ligands and the two axial ligands. The use of two distinct subsets of atoms for start and scope also allows the evaluation of different functions on the molecular graph such as the differences of atomic properties,

$$_{\text{scope}}^{\text{start}}P'_d = \sum_{i \in \text{start}} \sum_{j \in \text{scope}} (P_i - P_j)\delta(d_{ij}, d), \qquad (2)$$

which are evaluated for the mc/all and lc/lig start/scope combinations. Separate averages of the ligand charge and denticity over the equatorial and axial ligands complete the set of 155 nontrivial descriptors referred to as standard-RACs in this work.

Due to either the involvement of multiple ligands or their non-local character, most of the start/scope combinations of standard-RACs are incompatible with the concept of a many-body expansion localized at the metal center. Thus, we only use the $_{\text{lig}}^{\text{lc}}P_d$ and $_{\text{lig}}^{\text{lc}}P'_d$ sets to featurize the ligand dependence of the many-body interaction terms. As a consequence of this restriction to only ligand-connecting atom-centered descriptors, each connecting atom of a multidentate ligand is treated like a separate monodentate ligand. For closely related ligands such as bipyridine and pyridine, this treatment results in identical descriptors up to a depth of $d=2$ and a potential source of information loss for the ML models. Nevertheless, we choose not to add an additional feature to encode ligand denticity since, for all ligands in our data set, any ambiguity is resolved by the inclusion of descriptors for depth $d=3$. Finally, the ligand charge, or a fraction thereof in the case of multidentate ligands, e.g., -1/2 for a bidentate ligand with charge -1, is added as a feature yielding a vector $\mathbf{L}_i$ consisting of a total of 32 nontrivial features referred to as ligand-RACs throughout the remainder of this article. For both the standard-RACs and the ligand-RACs, the character of the metal ion is featurized using a vector $\mathbf{M}$ containing seven entries corresponding to separate one-hot encodings of the electron configuration ($d^3$, $d^4$, $d^5$, $d^6$, or $d^7$), and the oxidation state (II or III).

**3. Data sets.**

We curated a data set of octahedral TMCs from six of our previous studies on the prediction of electronic structure properties from empirical TMC featurizations[78], graph-based descriptors for inorganic chemistry[55], genetic algorithm optimization of spin crossover complexes[79], machine learning predictions of DFT geometry optimization outcomes[80], systematic enumeration of monodentate and bidentate ligand space[45], and a comparison of 3d and 4d TMCs[81]. For the present study, we selected complexes composed of four $3d$ metals in two oxidation states M(II)/M(III), where M = Cr, Mn, Fe, or Co. We further restricted our data set to complexes with computed DFT properties for both high-spin (HS) and low-spin (LS) states, where HS and LS are defined as quintet and singlet for both $d^4$ Mn(III)/Cr(II) and $d^6$ Co(III)/Fe(II), sextet and doublet for $d^5$ Fe(III)/Mn(II), and quartet and doublet for both $d^3$ Cr(III) and $d^7$ Co(II). From this initial data set of 2,350 TMCs, we removed 349 data points with positive highest occupied molecular orbital (HOMO) energies, which are common in complexes with a large number of negatively charged ligands. Following the protocol in prior studies[82], structures with spin expectation values $\langle S^2 \rangle$ deviating by more than 1.0 from the expected values (136 examples) and deviations from the expected octahedral geometry (59 examples) measured by previously established metrics[75,80,82] were eliminated (Supplementary Material Table S1 and Table S2). The final data set consists of 1,806 pairs of HS and LS complexes comprising 107 unique ligands (72 monodentate, 34 bidentate, 1 tetradentate).

We trained ML models to predict seven target properties of these TMCs: the adiabatic spin-splitting energy, $\Delta E_{\text{H-L}}$, i.e., the difference in energies between the geometry-optimized HS and LS states, four frontier orbital energies corresponding to the highest occupied and lowest unoccupied molecular orbital (HOMO and LUMO) energies for each of the spin states, and the HOMO–LUMO gaps of each spin state. In open-shell systems these molecular orbital energies can be defined for both the majority α and minority β spin. To resolve this ambiguity, we used a purely energy-based convention for HOMO and LUMO, i.e., the HOMO and LUMO levels are defined as $\max(\text{HOMO}_\alpha, \text{HOMO}_\beta)$ and $\min(\text{LUMO}_\alpha, \text{LUMO}_\beta)$, respectively.

In addition to the data set curated from previous studies, we generated two new test sets to measure the performance of models with respect to their ability to generalize from homoleptic data to heteroleptic complexes and to new ligands. The first set, the composition test set, is composed of all binary combinations and structural isomers for a given composition of three monodentate ligands: methanol, hydrogen cyanide, and hydrogen isocyanide. These ligands were selected because they cover a large range of field strengths in the spectrochemical series. The training set from the data curated from previous studies contains the homoleptic complexes for all investigated metal centers but no heteroleptic complexes built from the three ligands. Combining the eight different metal and oxidation state pairings, the eight compositional and structural isomers of binary complexes, and the three possible ligands pairings yields 192 TMCs, all of which passed the checks for octahedral geometry and spin expectation value. The second set, the ligand test set, is composed of homoleptic complexes of ligands not present in the training data from previous studies. We selected a total of 16 monodentate and 5 bidentate ligands by identifying unique chemical motifs in the 56 neutral, closed-shell ligands from the screening of the Cambridge Structural Database[83] for Fe(II) complexes in Ref. 47 and by selecting ligands from the enumeration of ligands in Ref. 45 that were not part of the original DFT-evaluated subset. Combining the 8 different metal and oxidation state pairings with the 21 new ligands yields 168 complexes, 127 of which passed the checks for octahedral geometry and spin expectation value (Supplementary Material Table S3).

The curated data set of previously studied complexes is split into a training data set and a validation set using a two-step approach. First, complexes containing any ligand belonging to a hand-selected set of 11 ligands representing the most common coordination environments are placed into the validation set (156 complexes, Supplementary Material Text S1). This ensures that the validation set metrics measure not only the performance on different combinations and structural isomers of the training set ligands but also the generalization to previously unseen ligands. In the second step, randomly selected examples are added to the validation set until a final data split of roughly 80/20 is achieved, yielding training and validation set sizes of 1444 and 362 complexes, respectively. All target values are normalized by subtracting their respective training set mean and dividing by the training set standard deviation. The RAC-type features are rescaled to the interval [-1, 1] by dividing each vector component by the largest absolute value observed in the training set. This approach is chosen because it conserves zero values in the feature vectors, which take a special role in RACs descriptors (typically encoding the absence of an atom), and results in a consistent feature range when combined with the one-hot encoded metal center features.

## 4. Computational details.

### 4.1 Electronic structure calculations.

DFT properties were obtained for the newly generated test sets (see Sec. 3) first by building the initial geometries using molSimplify[82,84]. DFT geometry optimizations were carried out using a developer version of the TeraChem[85-87] graphical processing unit accelerated quantum chemistry package. We use the B3LYP[88-90] hybrid functional combined with the LANL2DZ effective core potential[91] for iodine atoms and all transition metals and the 6-31G* basis[92,93] for the remaining atoms. This combination is chosen for consistency with the curated data set. However, we have previously shown a high linear correlation across different density functional approximations (DFAs) for several TMC properties.[94] This suggests that the presented findings should be independent of the specific choice of DFA. Singlet spin states were evaluated with the spin-restricted formalism, while all other calculations are performed in the spin-unrestricted formalism. To aid convergence, level shifting[95,96] of 0.25 hartree on both majority- and minority-spin virtual orbitals were employed for all calculations. Geometry optimizations are performed with the BFGS algorithm in translation-rotation-internal coordinates[97] using the default convergence criteria of 1 × 10$^{-6}$ hartree for the energy change between steps and 4.5 × 10$^{-4}$ hartree/bohr for the maximum gradient component.

### 4.2 Machine learning models.

In this work we use feed-forward artificial neural networks (NNs) and kernel ridge regression (KRR) to model the individual MBE interaction terms. However, the presented MBE-based approach is compatible with any ML regression algorithm. Regardless of the specific choice of ML approach, the three-body interactions involving dissimilar ligands need to be symmetrized with respect to the exchange of the two ligands, i.e, to ensure that $q_3(\mathbf{M}, \mathbf{L}_i, \mathbf{L}_j) = q_3(\mathbf{M}, \mathbf{L}_j, \mathbf{L}_i)$. We implement this permutational invariance by applying a feature-wise transformation, calculating the average $\bar{\mathbf{L}}_{ij} = ½ (\mathbf{L}_i + \mathbf{L}_j)$ and half the absolute difference $\tilde{\mathbf{L}}_{ij} = ½ |\mathbf{L}_i - \mathbf{L}_j|$ of the two ligand feature vectors. Alternatively, the symmetry condition could be realized by evaluating the ML interaction term for both ligand permutations and subsequently averaging. This is equivalent to lifting the restriction on the second sum over ligands in the three-body contribution of eq (1) and multiplying the term by a factor of one half. For models trained to predict the four frontier orbital energies, we employ multi-task learning throughout. In the ANN model, this is implemented using shared hidden layers between the four targets and separate linear output layers for each orbital energy. For the KRR model, the multi-task approach corresponds to a shared kernel matrix and, thereby, shared hyperparameters for the models of the four frontier orbital energies. This restricted architecture is motivated by a high correlation between the frontier orbital energies, i.e., the Pearson correlation coefficient is greater than 0.96 for all pairwise combinations (Supplementary Material Figure S1).

### 4.2.1 Artificial neural networks.

The implementation of the MBE-based NN models was simplified by automatic differentiation, a core feature of modern neural network libraries. In this work we use the Keras[98] and Tensorflow[99] Python packages. The individual interaction terms are modeled using fully-connected layers combined with the softplus activation function[100] for hidden layers and a linear activation on the output layer. To reduce overfitting we apply $L_2$ weight-regularization with strength $\lambda$ and dropout,[101,102] i.e., random zeroing of nodes during training with a probability $p$, on all hidden layers (Supplementary Material Figure S2).

The training was carried out using the Adam optimizer[103] and the mean squared error loss function for a maximum of 5000 epochs. Early stopping was performed by monitoring the mean squared error on the validation set and stopping the training if the error did not decrease for 100 epochs. Models containing three-body interaction terms were trained in three steps, each terminated by the early stopping procedure. First, a simplified model using only two-body terms was trained. In the second step, the three-body terms were fit as corrections to the simple model by keeping the weights for the two-body terms fixed. In the final training step, the weights of all interaction terms were allowed to vary without restrictions. The size of the NNs, the batch size, the regularization strength $\lambda$, and the dropout probability $p$ are determined by minimizing the mean absolute error on the validation set using the tree-structured Parzen estimator[104] approach with 100 evaluations as implemented in the HyperOpt[105] package (Supplementary Material Table S4 and Table S5). Using the optimized hyperparameters, we constructed an ensemble model by training 10 independent models from different random weight initializations. The presented ensemble predictions and uncertainties were evaluated by calculating the mean and standard deviation over predictions of the 10 independent models.

### 4.2.2 Kernel ridge regression.

It can be shown that incorporating the MBE approach into a kernel-based method such as KRR results in a modified kernel function (Appendix B). Given the additive nature of the MBE, this new kernel function consists of a linear combination of terms corresponding to each expansion order in the MBE,

$$k(\mathbf{x}, \mathbf{y}) = k^{\text{one-body}}(\mathbf{x}, \mathbf{y}) + k^{\text{two-body}}(\mathbf{x}, \mathbf{y}) + k^{\text{three-body}}(\mathbf{x}, \mathbf{y}) + \cdots, \qquad (4)$$

where the feature vectors $\mathbf{x}$ and $\mathbf{y}$ comprise the metal center featurization $\mathbf{M}$ and the six ligand featurizations $\mathbf{L}_i$, i.e., $\mathbf{x} = [\mathbf{M}_x, \mathbf{L}_{x,1}, \mathbf{L}_{x,2}, \mathbf{L}_{x,3}, \mathbf{L}_{x,4}, \mathbf{L}_{x,5}, \mathbf{L}_{x,6}]$. The one-body kernel term is a function of just the metal core featurizations $\mathbf{M}$,

$$k^{\text{one-body}}(\mathbf{x}, \mathbf{y}) = k_1(\mathbf{M}_x, \mathbf{M}_y), \qquad (5)$$

where $k_1$ is any valid kernel function. Motivated by the one-hot encoding used to featurize the metal core, we use a linear dot product kernel $k_1(\mathbf{M}_x, \mathbf{M}_y) = \mathbf{M}_x \bullet \mathbf{M}_y$ for the one-body kernel. The two-body kernel term is given by,

$$k^{\text{two-body}}(\mathbf{x}, \mathbf{y}) = \frac{c_2}{36} \sum_i^6 \sum_j^6 k_2\big([\mathbf{M}_x, \mathbf{L}_{x,i}], [\mathbf{M}_y, \mathbf{L}_{y,j}]\big), \qquad (6)$$

where $c_2$ is a scalar weighting factor and the kernel $k_2$ is a function of two vectors, each of which is a concatenation of the metal featurization and a single ligand featurization. The relative importance of the two-body contributions is tuned by the hyperparameter $c_2$. Additionally, a factor of 1/36 is used to rescale the kernel to unity $k^{\text{two-body}}(\mathbf{x}, \mathbf{x}) = 1$ for homoleptic complexes, $c_2=1$, and normalized kernel functions $k_2$. Note, however, that the two-body kernel itself is not normalized since $k^{\text{two-body}}(\mathbf{x}, \mathbf{x}) \neq 1$ for heteroleptic complexes. Finally, the three-body kernel term can be divided into *cis*-type and *trans*-type interactions,

$$k^{\text{three-body}}(\mathbf{x}, \mathbf{y})$$

$$= \frac{c_{3,cis}}{144} \sum_{i,j \in cis}^{12} \sum_{k,l \in cis}^{12} k_{3,cis}\left([\mathbf{M}_x, \bar{\mathbf{L}}_{x,ij}, \tilde{\mathbf{L}}_{x,ij}], [\mathbf{M}_y, \bar{\mathbf{L}}_{y,kl}, \tilde{\mathbf{L}}_{y,kl}]\right) \qquad (7)$$

$$+ \frac{c_{3,trans}}{9} \sum_{i,j \in trans}^{3} \sum_{k,l \in trans}^{3} k_{3,trans}\left([\mathbf{M}_x, \bar{\mathbf{L}}_{x,ij}, \tilde{\mathbf{L}}_{x,ij}], [\mathbf{M}_y, \bar{\mathbf{L}}_{y,kl}, \tilde{\mathbf{L}}_{y,kl}]\right),$$

where $\bar{L}_{ij}$ and $\tilde{L}_{ij}$ denote the symmetrized ligand features, $c_{3,cis}$ and $c_{3,trans}$ are scalar weighting factors, and $k_{3,cis}$ and $k_{3,trans}$ are kernel functions of the concatenation of the metal featurization and the two symmetrized ligand feature vectors. The scaling factors of $c_{3,cis}/144$ and $c_{3,trans}/9$ again tune the relative strength of the respective interaction terms and ensure normalization of the individual terms for homoleptic complexes and $c_{3,cis}=c_{3,trans}=1$. While it is difficult to assign a many-body order to standard-RACs ligand features, we follow a similar approach to construct a kernel function for the KRR model built from standard-RACs. The one-body kernel $k^{\text{one-body}}$, eq (5), is combined with a single kernel $k^{\text{RACs}}$ intended to capture all higher-order many-body contributions,

$$k(\mathbf{x}, \mathbf{y}) = k^{\text{one-body}}(\mathbf{x}, \mathbf{y}) + c_{\text{RACs}} k_{\text{RACs}}(\mathbf{x}, \mathbf{y}), \qquad (8)$$

where $c_{\text{RACs}}$ is a scalar weighting factor, and $k_{\text{RACs}}$ is a function of two vectors **x** and **y** obtained by concatenating the metal featurization with the standard-RACs ligand features.

The presented kernel modifications were implemented as an extension of the Gaussian process regression (GPR) module in scikit-learn[106] due to its higher degree of modularity compared to the KRR module. This also allowed us to obtain predictions of the KRR model uncertainty, as measured by the variance $\sigma^2$, by using the corresponding GPR expression,[107]

$$\mathbb{V}[Q_*^{\mathrm{ML}}] = \sigma^2(\mathbf{x}^*) = k(\mathbf{x}^*,\mathbf{x}^*) - \mathbf{k}^{*\mathsf{T}}(K + \lambda)^{-1}\mathbf{k}^*, \tag{9}$$

where $\mathbf{x}^*$ is the feature vector of a given complex, $k$ is the total kernel function (eq (4) or eq (8)), $\mathbf{k}^*$ is a vector obtained by evaluating the kernel function for $\mathbf{x}^*$ and all training examples, $K$ is the matrix of kernel values of pairwise combinations of training examples, and $\lambda$ is the regularization strength. Regardless of the reliance on GPR concepts and implementations, we refer to the presented approach as KRR because GPR usually also implies Bayesian selection of the hyperparameters by maximizing the log marginal likelihood, which we did not do.

We investigated two different choices for the kernel functions $k_{\mathrm{RACs}}$, $k_2$, $k_{3,cis}$, and $k_{3,trans}$ used in the ligand interaction terms, the radial basis function (RBF) kernel,

$$k^{\mathrm{RBF}}(\mathbf{x},\mathbf{y}) = \exp\left(-\frac{\|\mathbf{x}-\mathbf{y}\|^2}{2l^2}\right), \tag{10}$$

and the Matérn kernel (for $\nu = 3/2$), [108,109]

$$k^{\mathrm{Matérn}}(\mathbf{x},\mathbf{y}) = \left(1 + \frac{\sqrt{3}}{l}\|\mathbf{x}-\mathbf{y}\|\right)\exp\left(-\frac{\sqrt{3}}{l}\|\mathbf{x}-\mathbf{y}\|\right), \tag{11}$$

where, in both cases, $l$ is a hyperparameter typically referred to as the kernel length scale. Analogously to the neural network approach, the KRR hyperparameters, i.e., the regularization strength $\lambda$, the choice of kernels $k_{\mathrm{RACs}}$, $k_2$, $k_{3,cis}$, and $k_{3,trans}$, and their respective weighting factors $c$ and length scales $l$, are determined by minimizing the mean absolute error on the validation set using 100 iterations of the tree-structured Parzen estimator approach. To reduce the hyperparameter search space for the three-body model, all hyperparameters of its two-body kernel are fixed to the values obtained during the hyperparameter search for the two-body model (Supplementary Material Table S6 and Table S7).

## 5. Results and discussion.

### 5.1 Machine learning prediction of spin-splitting energies.

We first evaluate the fitting of the training set for the newly derived models using two-body and three-body terms in combination with KRR and NN models on the spin-splitting energy and compare them to the standard-RACs based approach. The mean absolute errors (MAEs) of the many-body ML model predictions show similar trends with respect to the featurization for both the KRR and the NN approaches (Table 1). When trained on the standard-RACs features, both of the ML methods achieve the lowest training set errors, of 0.9 kcal/mol and 2.3 kcal/mol for KRR and NN, respectively. The models based on the MBE up to two-body order are less expressive and, therefore, result in significantly higher MAEs on the training set. This limitation is lifted by the inclusion of three-body terms. The corresponding three-body models yield slightly larger training set MAEs than the best-performing standard-RACs models. We also observe significantly lower training errors for all three KRR-based models compared to their NN counterparts. This can be attributed to the fact that the KRR method results in perfect interpolation in the limit of zero regularization, under the assumption of a unique featurization. Neural networks on the other hand are also limited in accuracy by the finite-sized architecture and the stochastic training process in addition to the $L_2$ regularization on their weights.

**Table 1**. Mean absolute errors of the model predictions of the spin-splitting energies on all four data sets in kcal/mol. The lowest error for each data set is indicated in bold.

| Model | Training set | Validation set | Composition test set | Ligand test set |
|---|---|---|---|---|
| KRR standard-RACs | **0.87** | 3.67 | 4.10 | 4.84 |
| KRR two-body | 2.63 | 3.96 | 3.12 | 5.05 |
| KRR three-body | 1.03 | **3.30** | **2.75** | 4.93 |
| NN standard-RACs | 2.33 | 3.51 | 4.16 | 4.73 |
| NN two-body | 3.16 | 3.73 | 3.61 | 4.14 |
| NN three-body | 2.78 | 3.48 | 3.41 | **4.00** |

While fitting to training data gives a sense of how well the model is capturing the data, the comparison of model performance can only be fairly assessed on errors for the validation set. On the validation set, the variation in MAEs between different featurization concepts and ML approaches is significantly smaller than on the training set. Nevertheless, we generally observe the trend that all validation MAEs are higher at around 3.0–3.5 kcal/mol, and the standard-RACs based models are no longer the best-performing models (Table 1). While there is no clear difference in performance between the two ML methods, both approaches exhibit the same relative ordering with respect to the featurization. The three-body MBE-based models achieve the lowest MAEs of 3.3 kcal/mol and 3.5 kcal/mol for the KRR and NN models, respectively (Table 1). The models based on the MBE truncated at the two-body order again exhibit the highest MAEs of 4.0 kcal/mol for the KRR model and 3.7 kcal/mol for the NN model. We attribute this similarity of MAEs across all methods to the fact that the MAE on the validation set was used as target of the hyperparameter optimization. Because both models trained on standard-RACs result in slightly larger validation errors and exhibit lower training set errors, this hints at a higher degree of overfitting. Common to all six models is a weak dependence of the error on the identity and oxidation state of the central metal ion (Supplementary Material Table S8). With the exception of the KRR model using standard-RACs, the lowest MAEs are observed on Cr(III) complexes. We attribute this to the relatively low variance in the spin-splitting energies on the Cr(III) subset (Supplementary Material Figure S3). This is corroborated by significantly lower $R^2$ scores of between 0.2 to 0.3 on the Cr(III) subset compared to $R^2$ scores of approximately 0.96 on the full validation set (Supplementary Material Table S9 and Table S10). Conversely, the highest MAEs are obtained on the Co(III) and Fe(II) subsets, which exhibit the highest range in target values.

The distribution of errors on the validation set highlights several outliers that are shared among the different ML method and featurization approaches (Figure 1 and Supplementary Material Figure S4). The standard-RACs KRR model exhibits the highest deviation from the DFT reference on Co(III)(pyr)$_4$(CN)(CH$_2$O) (where pyr = pyridine) for which the spin-splitting energy is underestimated by 23.4 kcal/mol (Figure 1a). Using the kernel function $k$ from eq. 8 as distance metric in the standard-RACs feature space, we identify the three complexes from the training set that are most similar to this outlier, Co(III)(pyr)$_4$(CO)$_2$, Co(III)(pyr)$_4$(H$_2$O)(CO), and Co(III)(pyr)$_4$(CO)(misc) (where misc = methylisocyanide), all of which have a spin-splitting energy 10–20 kcal/mol lower than Co(III)(pyr)$_4$(CN)(CH$_2$O). Therefore, we attribute the large error on this complex to the fact that the standard-RACs feature space distance is dominated by the four axial pyridine ligands and does not sufficiently account for the presence of cyanide, one of the strongest ligands in our data sets. This hypothesis is corroborated by the fact that Co(III)(pyr)$_4$(CN)(CH$_2$O) also represents the second-largest outlier for the standard-RACs NN model with an underestimation of 19.4 kcal/mol, while the MBE based models show significantly lower errors on this complex. Another commonality is observed for the three NN models that exhibit their highest errors, an overestimation of roughly 20 kcal/mol, on the complex Fe(III)(CHPH$_2$)$_6$ (Figure 1b). The only occurrence of this ligand in the training set is in combination with the ion Cr(III). Since the spin-splitting energy of Cr(III) is largely independent of the ligands, the ligand strength of CHPH$_2$ cannot be inferred from this complex (Supplementary Material Figure S3). The fact that the NN models predict significantly higher spin-splitting energies for Fe(III)(CHPH$_2$)$_6$ than the KRR models is attributed to the different extrapolation behavior of the two ML methods.

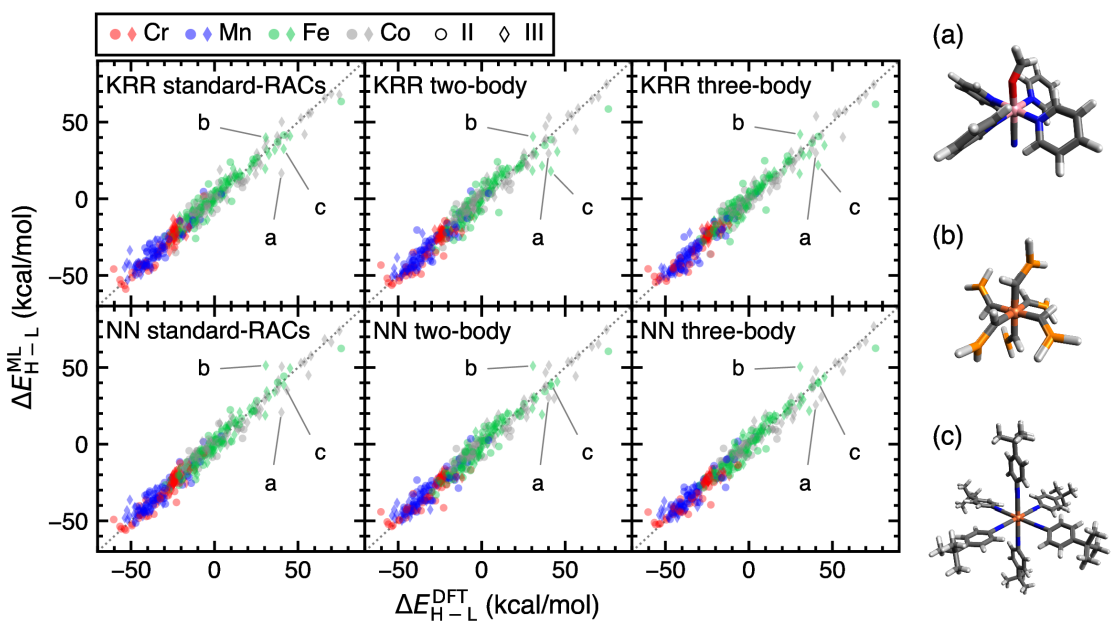

**Figure 1**. Calculated versus predicted spin-splitting energies of the validation set complexes for the two ML methods (rows) and the three featurization approaches (columns) all in kcal/mol. The three complexes with the largest prediction errors, Co(III)(pyridine)$_4$(cyanide)(formaldehyde) (a), Fe(III)(CHPH$_2$)$_6$ (b), and Fe(III)(pisc)$_6$ (c), are annotated in all panels and their respective low-spin structures are depicted on the right. Structures are colored as follows: pink for Co, brown for Fe, gray for C, blue for N, white for H, red for O, and orange for P.

As a final example of shared outliers, both MBE-based KRR models exhibit the highest deviation from the DFT reference on the heteroleptic 1-tert-butyl-4-phenyl-isocyanide (pisc) Fe(III) complex. We attribute this to the fact that the training set does not contain any combination of pisc ligands and Fe(III) (Figure 1c). This results in two potential ways a ML model might leverage training data to inform prediction of the spin-splitting energy of Fe(III)(pisc)$_6$. One option is to construct the prediction from the metal–ligand interaction of Fe(III) with similar ligands. The two MBE-based KRR models follow this approach, as indicated by the fact that the three most similar training complexes as measured by their respective kernels are Fe(III)(CNH)$_6$, Fe(III)(misc)$_6$, and Fe(III)(misc)$_5$(CO). Alternatively, the ligand strength could be inferred from the interaction of pisc with other metal ions. Exemplary of this, the closest complexes in the training set as measured by the standard-RACs kernel are Mn(II)(pisc)$_6$, Fe(II)(pisc)$_6$, and Mn(III)(ox)$_4$(pisc)$_2$, which results in a significantly lower prediction error for the standard-RACs KRR model.

On the composition test set, which consists of all binary combinations and structural isomers of a fixed ligand set (see Sec. 3), we observe similar trends with regard to the featurization approach for both ML methods. That is, the three-body models achieve the lowest errors, especially in the KRR models (i.e., 2.8 kcal/mol vs. 4.1 kcal/mol for standard RACs), while two-body models exhibit intermediate performance (Table 1). We attribute the better performance of KRR over NN models to lower KRR errors on the homoleptic complexes in the training data set that are reference values for interpolation of the heteroleptic complexes. Despite these overall trends, outliers are evident when separately analyzing each of the eight metal/oxidation state combinations (Supplementary Material Table S11). On the Mn(III) subset, the global trend with regard to the featurization approach is reversed, and the standard-RACs models consistently outperform their MBE-based counterparts. Conversely, for the Fe(II) subset in combination with the three-body models, the highest $R^2$ scores (i.e., 0.98–0.99) are achieved (Supplementary Material Table S12).

To gain a better understanding of these trends, we analyze the interpolation behavior of the three KRR models on the Mn(III) and Fe(II) subsets in detail (Figure 2). The standard-RACs based model predictions show unsystematic deviations from an overall linear interpolation trend in the generalization to heteroleptic complexes. This results in potentially large differences of the predicted spin-splitting energy for the *cis*/*trans* and *fac*/*mer* isomers of the same composition. We attribute the overall linear trend in the predictions to the design of standard-RACs. In the RACs feature space, the heteroleptic complexes lie close to straight-line connections between their homoleptic counterparts (Supplementary Material Figure S5). In the two-body model, linear additivity is strictly enforced, resulting in a roughly 2 kcal/mol lower MAE on the Fe(II) subset (Supplementary Material Table S11). However, due to this simplification, the two-body model is unable to distinguish between the two isomers for a given composition. Furthermore, the limited model expressivity in the two-body model results in higher errors on the homoleptic complexes, which we attribute to a trade-off between homoleptic and heteroleptic complexes during model training. These shortcomings are addressed by the inclusion of three-body interaction terms, which introduce two additional degrees of freedom to the interpolation. Their effect can be described as curvature, i.e, systematic higher or lower predictions compared to the linear interpolation, and a splitting in the predicted spin-splitting energy of the *cis*/*trans* and *fac*/*mer* isomers (Appendix A). Both the KRR and NN three-body models predict small isomer splittings with a mean absolute deviation of roughly 0.6 kcal/mol between all isomer pairings in the composition test set. This is consistent with the DFT training data, which shows small energy differences and no clear pattern in the relative spin-splitting energy ordering of the isomers. Therefore, the curvature parameter is the main factor in the decrease in MAEs due to the three-body interaction terms. Comparison with the two-body interpolation curves shows that this additional degree of freedom allows these models to achieve high accuracy on heteroleptic complexes without simultaneously increasing the errors on the homoleptic complexes. However, an incorrect prediction of the sign of the curvature, as exemplified in the three-body KRR predictions along the interpolation curve between $Mn(III)(CNH)_6$ and $Mn(III)(MeOH)_6$ (where MeOH = methanol), results in increased MAEs compared to linear interpolation (Figure 2). A similar analysis on the other metal ion subsets and the NN models confirms that this is the source for the observed increase in MAEs for the three-body models over their corresponding two-body model on certain subsets (Supplementary Material Figures S6 and S7).

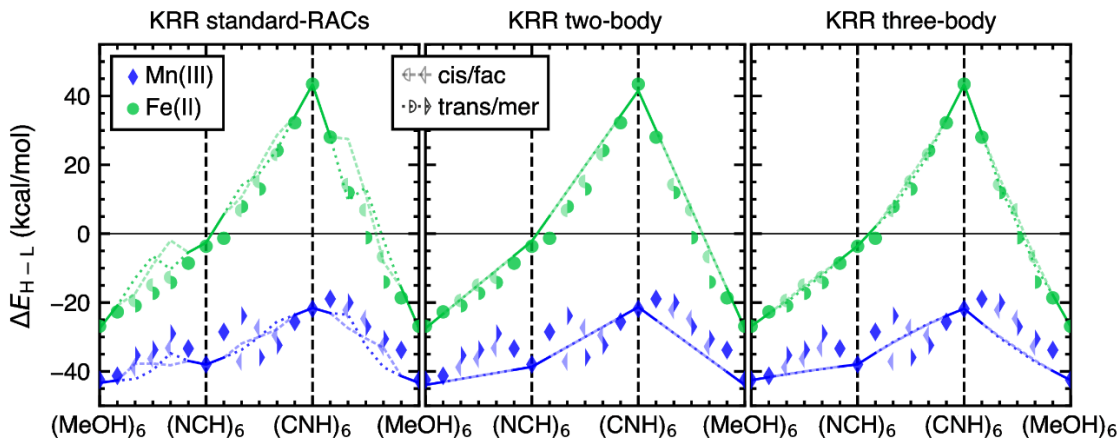

**Figure 2**. Plot of the spin-splitting energy (in kcal/mol) interpolation curves for binary complexes of three ligands, methanol (MeOH), hydrogen cyanide (NCH), and hydrogen isocyanide (CNH), and two metals Mn(III) and Fe(II) showing the DFT reference values (scatter) and the corresponding ML predictions (lines) for KRR models using different featurization approaches (columns). For compositions with multiple structural isomers, values are plotted using half markers for the DFT reference and dashed/dotted lines for the ML predictions.

Finally, we compare the generalization of the models to unseen ligands using the ligand test set. As expected, all models and featurizations yield slightly worse MAEs than for the validation set. All three KRR models show comparable performance, with the lowest MAE of 4.8 kcal/mol achieved by the standard-RACs models (Table 1). However, both MBE-based NN models exhibit significantly lower generalization errors of 4.1 kcal/mol for the two-body model and 4.0 kcal/mol for the three-body model. While the NN model on standard-RACs results in a slightly lower MAE of 4.7 kcal/mol than that for the KRR model, the standard-RACs featurization has the worst performance of the three NN models. Analysis of subsets for the eight metal/oxidation state combinations reveals that the KRR models exhibit significantly higher MAEs on the Cr(III) subset compared to the NN models (Supplementary Material Table S13). Given the low variation of spin-splitting energies on the Cr(III) subset, using the average value of the Cr(III) training set complexes as constant prediction results in a lower MAE of 2.5 kcal/mol than the KRR models with MAEs of roughly 4 to 5 kcal/mol. We attribute this to the fact that the kernel-based approaches insufficiently separate the effect of the metal identity from the effect of the ligands.

Addressing this issue would require a redesign of the kernel, for example, following the approach of Jørgensen in which the two-body interaction terms are modeled as a product of separate interaction strength functions for the metal and the ligand $q_2(M, L_i) = g(M)f(L_i)$.[60,110] With the exception of the Co(III) subset, the overall dependence of the model MAEs on the metal/oxidation state combinations for the ligand test set is consistent with results from the validation and composition test sets. Dividing the ligand test set into subsets corresponding to the 21 different ligands shows significantly higher MAEs for the subset of the ligands thiane and PHCHOH, which also represent outliers in an analysis of the feature space distributions (Supplementary Material Table S14 and Figure S5). Removal of the 11 complexes corresponding to these two ligands results in a decrease in the MAE of roughly 0.5 kcal/mol for all models while maintaining the significantly improved generalization errors for the MBE-based NN models.

For applications in chemical discovery, large prediction errors on certain complexes do not represent a problem as long as the ML models are able to correctly identify these as outliers. The NN ensemble approach allows one to evaluate the prediction uncertainty as measured by the standard deviation of the ten individual neural networks.[111,112] Both the standard-RACs and the two-body NN models severely underestimate the prediction uncertainty even on significant outliers (Figure 3). We attribute this to the limited expressivity of these models caused by their respective featurization approaches that limits variation in parameters of the NN models in the ensemble, which is less severe for the three-body model that correctly assigns high uncertainty to the thiane complexes. We evaluate the average negative log-likelihood (NLL) as a quantitative measure for this model uncertainty prediction, for which the three-body NN exhibits up to a three times lower average NLL score on all test sets than the two-body and standard-RACs NN models (Supplementary Material Table S15). An analysis of the uncertainty prediction for KRR shows that the average predicted standard deviation on the ligand test set (calculated with equation 9) ranges from 40 kcal/mol for the standard-RACs model to 160 kcal/mol for the three-body model, which produces low NLL scores but makes these models' predicted uncertainties too high to be useful in Bayesian optimization algorithms (Supplementary Material Table S15). We attribute the large predicted uncertainties to our optimization of KRR hyperparameters using the validation set MAE instead of a fully Bayesian approach, which was done to ensure fair comparison to the NN results.

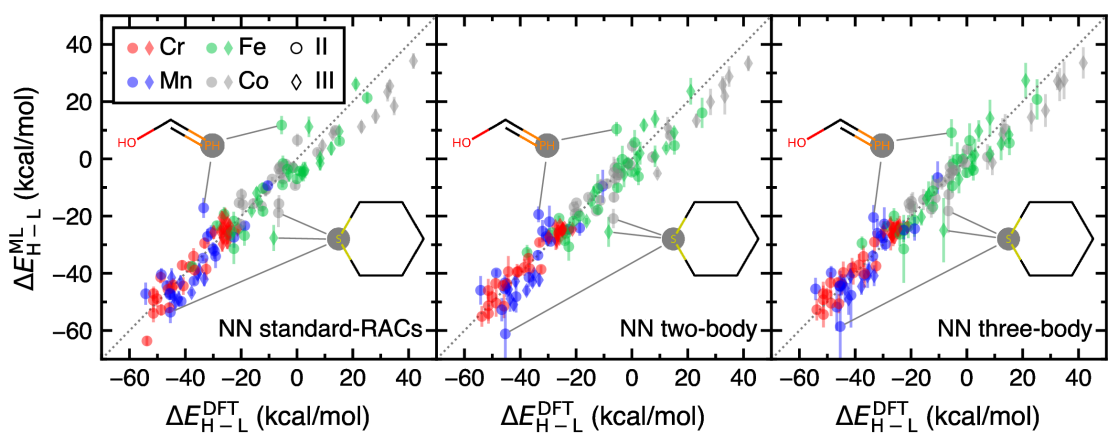

**Figure 3**. Calculated versus NN-predicted spin-splitting energies of the ligand test set for the three featurization approaches (columns). Error bars corresponding to two standard deviations depict the model uncertainty. The five complexes with the largest prediction errors, Fe(III)(thiane)$_6$, Mn(II)(thiane)$_6$, Co(II)(thiane)$_6$, Fe(II)(PHCHOH)$_6$, and Mn(II)(PHCHOH)$_6$, are annotated in all panels using a 2D representation of the respective ligand in which the connecting atom is highlighted with a gray circle.

**5.2 Machine learning prediction of frontier orbital energies.**

To test the generalization of our suggested MBE ML approaches, we next considered the frontier orbital energies, which are less metal-centric properties, as a second class of target properties to evaluate the performance. We might expect the ML model accuracy on both the composition test set and the ligand test set to be significantly improved by the inclusion of the MBE over standard RACs. However, the spin-splitting energy represents a best-case scenario for these models because the MBE of this property is closely related to the concept of the spectrochemical series, whereas the frontier orbital energies are more sensitive reporters of the overall size and charge of the molecule. Of the frontier orbital energies, we first focus our analysis on the HOMO energies, as we have previously successfully applied additivity relations to predict the HOMO energies of heteroleptic singlet Fe(II) complexes from their homoleptic counterparts.[47]

Given the high degree of linear correlation between the LS and HS HOMO energies, we do not distinguish between the two spin-multiplicities in our analysis (Supplementary Material Figure S1). The presence of negatively charged ligands in our data sets, which shift the HOMO level towards higher energies, produces a large variance of the target data. In the training data set, the HOMO energies range from roughly -25 eV to 0 eV (Supplementary Material Figure S8). For all four metal ions, the M(III) subset covers a larger range of values than the M(II) counterparts (average range of 25 eV for M(III) vs. 16 eV for M(II)). We attribute this to the fact that, on average, the median M(III) HOMO energies are 4.2 eV lower than the corresponding M(II) energies. This lower median HOMO level in turn allows complexes with a larger total number of negative charges on the ligands to pass the check for negative spin-majority HOMO energies applied during the data set curation (see computational details in Sec. 4). This effect is the result of more than just compensating for the difference in charge of metal ion because the M(III) subsets contain more overall negatively charged complexes (Supplementary Material Table S16). Given the large variance in the target data due to charged ligands and the fact that this property is less metal-centric, we would expect that learning the frontier orbital energies to be significantly more challenging than the spin-splitting energy.

Consistent with observations for spin-splitting energies, the KRR models outperform the NN counterparts on the training set. The KRR models have MAEs of 0.15 eV for the standard-RACs model (vs. 0.29 eV for the NN), 0.41 eV (vs. 0.43 eV for the NN) for the two-body model, and 0.22 eV for the three-body model (vs. 0.32 eV for the NN, Table 2). On the validation set, the difference in performance between the different ML methods and featurization approaches is again significantly smaller. We observe comparable MAEs of roughly 0.4 eV for all models, where the NN models exhibit on average 0.02 eV lower errors than the KRR models (Table 2). Given the large range of target values, the MAEs on the training and validation set correspond to high $R^2$ scores of over 0.97 for all models (Supplementary Material Table S17).

**Table 2.** Mean absolute errors of the model predictions of the combined LS and HS HOMO energies on all four data sets in eV. The lowest error for each set is indicated in bold.

| Model | Training set | Validation set | Composition test set | Ligand test set |
|---|---|---|---|---|
| KRR standard-RACs | **0.15** | 0.37 | 0.58 | 1.03 |
| KRR two-body | 0.41 | 0.44 | 0.28 | 1.27 |
| KRR three-body | 0.22 | 0.40 | **0.23** | 1.12 |
| NN standard-RACs | 0.29 | **0.34** | 0.56 | **0.96** |
| NN two-body | 0.43 | 0.43 | 0.30 | 1.05 |
| NN three-body | 0.32 | 0.38 | 0.79 | 0.97 |

The composition test exhibits the highest difference in MAEs between the models (Table 2). For the KRR models, we observe the expected trend of a significant improvement by including the MBE resulting in a MAE of 0.28 eV for the two-body model, almost half the 0.58 eV MAE of the standard-RACs model. The NN models show a similar improved accuracy of the two-body model, with a MAE of 0.30 eV, over the standard-RACs model, with a MAE of 0.56 eV. However, while the three-body KRR model achieves the overall lowest MAE on the composition test set of 0.23 eV, the three-body NN model exhibits the highest MAE of 0.79 eV (see more discussion, next). Finally, on the ligand test set, we observe high errors for all models, where the standard-RACs NN achieves the lowest MAE of 0.96 eV and the two-body KRR model exhibits the highest MAE of 1.27 eV. This poor performance of all models on the generalization to previously unseen ligands again highlights the fact that the HOMO energies are a less metal-centric and, therefore, a more challenging property to predict when prior knowledge of a specific ligand chemistry is unavailable to the model.

The uncharacteristically high MAE of the three-body NN on the composition test set warrants a closer analysis. As discussed previously, the effect of the three-body interactions on binary heteroleptic complexes can be described in terms of an isomer splitting, i.e., the difference in predictions for *cis*/*trans* and *fac*/*mer* isomers, and "curvature", i.e., the deviation from linear interpolation for the 5+1 complexes (Appendix A). Averaging over the three ligand combinations and eight metal/oxidation state combinations of the composition test set yields an average curvature of 0.07 eV for the three-body KRR model and 0.55 eV for the three-body NN model. The large predicted curvature of the three-body NN results in increasing MAEs along the interpolation curve of 0.56 eV for the $M(L_A)_5(L_B)_1$ complexes, 0.83 eV for the $M(L_A)_4(L_B)_2$ complexes, and 0.93 eV for the $M(L_A)_3(L_B)_3$ complexes (Figure 4 and Supplementary Material Table S18). We interpret this as overfitting of the stepwise change in HOMO levels observed on the M(III) complexes $M(L_A)_5(L_B)_1$ containing a methanol ligand (Supplementary Material Figure S9). This finding is further compounded by the fact that the training and validation set do not contain any *fac* or *mer* complexes, and high errors on these compositions, therefore, do not influence the training and hyperparameter optimization procedures. However, overfitting alone does not explain why the same limitations are not observed for the three-body KRR model (Supplementary Material Figure S10). Another contributing factor might, therefore, be the iterative training approach used for the three-body NN, in which the three-body terms are fitted as a correction to a two-body model (see Sec. 4.2.1). That is, in the second training phase, the two-body terms are fixed and the three-body interactions are fitted on the residuals from the first training phase. This most likely results in an overcorrection that is not properly resolved during the final phase of training in which all terms are fitted together. In summary, the comparison of the two different three-body models on the composition test set shows that the additional flexibility that allows the three-body KRR to outperform its two-body counterpart instead resulted in overfitting of the three-body NN due to our specific training algorithm and insufficient training data.

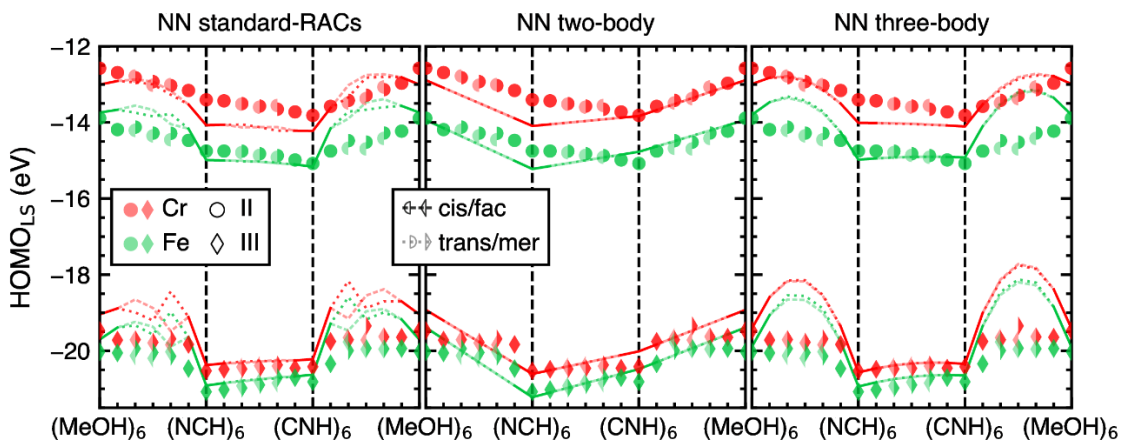

**Figure 4**. Plot of the LS HOMO energy (in eV) interpolation curves for binary complexes of three ligands, methanol (MeOH), hydrogen cyanide (NCH), and hydrogen isocyanide (CNH), and two metals Cr and Fe showing the DFT reference values (scatter) and the corresponding ML predictions (lines) for NN models using different featurization approaches (columns). For compositions with multiple structural isomers, values are plotted using half markers for the DFT reference and dashed/dotted lines for the ML predictions.

We next considered the performance of the different ML methods and featurization approaches on the more stringent ligand test set. On average, each approach produces comparably worsened MAEs (ca. 0.9–1.1 eV), but there are some smaller differences (Table 2). Similar to our finding on the spin-splitting energies, we observe that for all three featurization approaches the NN models yield lower MAEs than the KRR models in this extrapolative regime. Consistent with the results on the other data sets, all models achieve significantly lower MAEs on the M(II) subsets of the ligand test set than on the M(III) subsets (on average over 0.4 eV lower, Supplementary

Material Table S19). We attribute this to the fact that the M(III) portion of the training set covers a wider range of ligand charges, resulting in a more challenging generalization task. Analyzing the MAEs on subsets grouped by each of the 21 ligands (i.e., with varying metal or oxidation state), shows large variations ranging from 0.09 eV for the three-body KRR model on the M(DMF)$_6$ (where DMF = dimethylformamide) subset to 2.98 eV for the two-body KRR model on the M(oxz)$_6$ (where oxz = oxazoline) subset (Supplementary Material Table S20). One of the few examples of a ligand subset on which all models underestimate the HOMO energies is the aforementioned set of N-coordinating oxz complexes. However, unlike for the spin-splitting energies, these trends rarely hold for all ML methods and featurization approaches, making it challenging to conclude any relationship among featurization, chemistry, and model performance. A final notable difference between the different models is their ability to capture the metal/oxidation state dependence and ligand dependence of the HOMO energies. For example, both the standard-RACs NN and the three-body NN show comparable MAEs on the M(oxz)$_6$ subset of 1.9 eV and 2.1 eV, respectively. However, for the three-body NN, this error stems from a systematic offset that is almost constant for all metal and oxidation state combinations, while the errors of the standard-RACs NN vary from 0.9 eV to 2.7 eV (Supplementary Material Figure S11). A quantitative measure of this is given by the standard deviations of the absolute error, which on this subset evaluate to 0.63 eV for the standard-RACs NN and 0.27 eV for the three-body NN (Supplementary Material Table S21). With the exception of a lower composition test set MAE for the three-body NN, all the trends discussed for the HOMO energies similarly hold for the LUMO energies (Supplementary Material Text S2, Tables S22 and S23, Figures S12 and S13). To summarize, the overall lack of trends for the MAEs on various subsets of the data and between the different ML methods and featurization approaches suggests that both the ligand-RACs and standard-RACs feature spaces do not sufficiently capture the similarity of ligands for the frontier orbital energies, resulting in high generalization errors. More data could address some of these limitations of all models, but the lack of trends between different models makes it challenging to identify what way the feature sets should be further improved.

An even more relevant property than the individual HOMO or LUMO energies is the difference between HOMO and LUMO levels, i.e., the HOMO–LUMO gap. Calculating this difference results in significant changes to the distribution of the DFT target values compared to the pure frontier orbital energies (Supplementary Material Figure S14). First, the high degree of linear correlation between the HOMO and LUMO energies of the same spin state results in cancellation of parts of the ligand dependence and, therefore, a smaller range of target values of about 7 eV. Second, the bimodal character of the frontier orbital distributions, stemming from the presence of charged ligands, cancels out almost completely. Third, the linear correlation between the LS and HS target values for the gap is significantly lower than for the individual HOMO and LUMO energies (Supplementary Material Figure S15 and Table S24). This can be attributed to cancellation of the linear correlation between the HOMO and LUMO energies of different spin states (Supplementary Material Figure S1). Finally, by definition, i.e., due to the Aufbau principle, the DFT reference values for the gap are strictly positive. We first examined ML-model predicted gaps obtained from independent model predictions of the HOMO and the LUMO. However, since the HOMO and LUMO energies are two largely independent model predictions, the described differences in the distribution of gap energies are not necessarily reflected in the ML predictions of the gap. For example, on a small number of complexes, the ML models predict a lower LUMO energy than HOMO energy, resulting in a negative predicted gap (Supplementary Material Table S25).

**Table 3.** Mean absolute errors of the model predictions of the combined LS and HS HOMO–LUMO gap energies on all four data sets in eV. The lowest error for each set is indicated in bold.

| Model | Training set | Validation set | Composition test set | Ligand test set |
|---|---|---|---|---|
| KRR standard-RACs | **0.16** | 0.52 | 0.45 | 0.80 |
| KRR two-body | 0.35 | 0.45 | 0.34 | 0.72 |
| KRR three-body | 0.20 | 0.41 | **0.26** | 0.61 |
| NN standard-RACs | 0.34 | 0.43 | 0.50 | 0.63 |
| NN two-body | 0.39 | 0.44 | 0.42 | 0.54 |
| NN three-body | 0.34 | **0.39** | 0.94 | **0.53** |

As expected, the MAEs of the gap predictions for all six models follow the same overall trends as the results on the HOMO and LUMO energies (Table 3). In fact, with the exception of the standard-RACs models, the MAEs on the training and validation set gap energies are within 0.01 to 0.04 eV of the MAEs on the HOMO energies and 0.08 to 0.12 eV higher than the MAEs on the LUMO energies. However, given the smaller range of the target values, these MAEs correspond to lower $R^2$ scores than for the HOMO or LUMO model predictions (Supplementary Material Table S26). On the composition test set, we observe that both standard-RACs models achieve slightly lower MAEs for the gap than on the frontier orbital energies, while all MBE-based models yield slightly higher MAEs. The significant outlier in the HOMO MAEs of 0.79 eV for the three-body NN on the composition set propagates into the predictions for the gap and results in an MAE of 0.94 eV, which we interpret as compounding of the errors on the HOMO and LUMO energies. Conversely, on the ligand test set, we observe cancelation of errors resulting in 0.2 to 0.5 eV lower MAEs compared to the HOMO energies. It is again important to note that these decreased MAEs do not correspond to higher $R^2$ scores due to the lower range of target values (Supplementary Material Table S26). In order to quantify the influence of the indirect modeling approach on ML performance, we trained additional models directly on the LS and HS gap energies as a point of comparison. This results in lower validation set MAEs for all models and significant reductions in both test set MAEs of the standard-RACs and two-body KRR models (Supplementary Material Table S27). On the other hand, these direct models exhibit an increased tendency to overfit, as evidenced by higher MAEs for the two three-body models on both test sets. We therefore interpret the approach of calculating the gap as a difference of the frontier orbital predictions as another regularization technique that might limit the achievable accuracy for already heavily regularized models but becomes crucial to reduce overfitting in highly expressive models.

To further investigate the differences between the frontier orbital models and the spin-splitting models, we evaluated the feature importance of the respective two-body NN models on the ligand test set using the model-agnostic Kernel Shapley Additive exPlanations (SHAP) approach.[113] As expected, both metal-centered features, d-electron configuration and oxidation state, rank in the top 5 most important features for all three properties (Figure 5). In fact, many of the other top 12 features are shared between the models. A notable difference is the fact that the spin-splitting model mostly relies on metal-local features, i.e., ligand RACs with depth $d \leq 2$, whereas for the prediction of the HOMO–LUMO gaps ligand RACs with the maximum depth considered in the feature set (i.e., $d = 3$) also play a significant role. This is in line with prior feature importance analysis across broader sets of transition metal complexes.[55,94] It is also consistent with our expectation that frontier orbital energies are a less metal-centric property than spin-splitting energies and, therefore, represent a more challenging target for all presented TMC models.

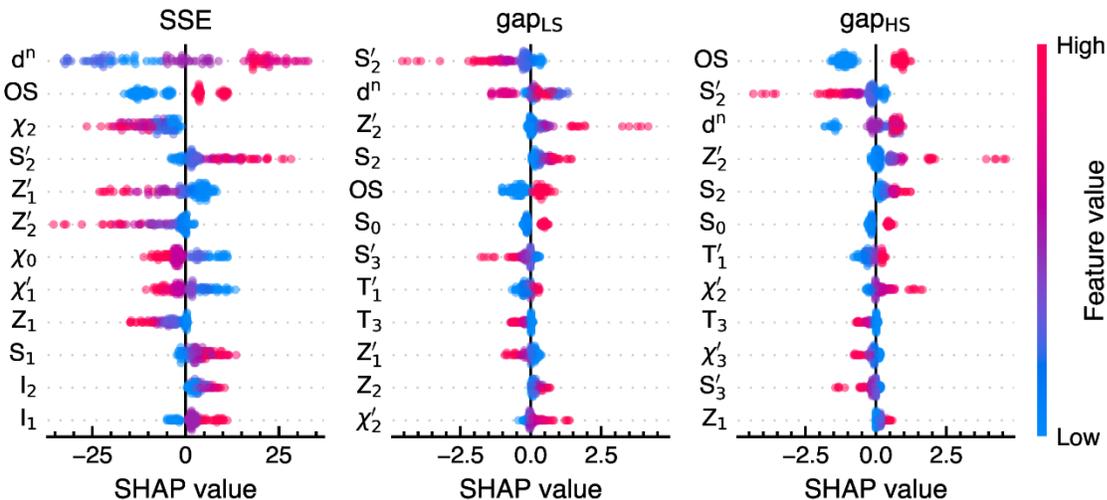

**Figure 5**. Kernel SHAP feature importance analysis of the two-body NN models for the spin-splitting energy (SSE, left), and the LS (center) and HS (right) HOMO–LUMO gaps (gap). For each model the 12 most important features, as measured by the mean absolute SHAP value, are shown, where $d^n$ refers to the d-electron configuration, OS to the oxidation state, and $P_d$ and $P_d'$ represent the ligand-centered product and difference RACs from equations 2 and 3. The model baseline is evaluated using a k-means clustering of the training set with k=25 and SHAP feature importance values are calculated for all 127 TMCs in the ligand test set. The one-hot encoded metal ion features are combined into two separate integer features ($d^n$ and OS) and their respective SHAP values are summed. Similarly, the six different starting atoms for the ligand-centered RACs are combined into a single feature and their SHAP values are summed.

In summary, the results for model performance on the frontier orbital energies in large parts mirror our findings on the spin-splitting energies. The KRR-based models outperform the their NN counterparts in the interpolative regime, i.e., on the composition test set, while the NN models show better generalization to previously unseen ligands. The inclusion of the MBE shows a systematic improvement of the accuracy on the composition test set for all frontier orbital energy targets. The only outlier from this trend, the three-body NN, highlights the increased risk of overfitting that comes with more flexible models. The standard-RACs based models achieved the lowest MAEs on the ligand test set for both HOMO and LUMO energies. However, for the HOMO–LUMO gaps, the MBE-based featurization approach results in improved error cancellation. The same systematic reduction of the MAEs with increasing MBE-order was observed on the composition test set.

**6. Conclusion.**

We introduced a novel approach based on the many-body expansion to encode the stereochemistry of octahedral TMCs in molecular graph descriptors. This featurization approach enables more efficient virtual high-throughput screening for material properties using ML surrogate models by exploiting ligand additivity relations. We derived the necessary expressions to implement this approach in two commonly used ML methods, KRR models and NNs. The MBE-based approach allows for the tuning of the expressivity of the ML models by truncating the expansion at different interaction orders. ML models obtained by truncation at the two-body and three-body order were investigated, and their performance was compared to models based on standard-RACs, our previous approach for featurization of octahedral TMCs.

We trained these models on a data set of 1,444 spin-splitting energies and observed improved model accuracy on test sets in the interpolative regime, i.e., TMCs composed of new combinations of ligands from the training set, and the extrapolative regime, i.e., TMCs consisting of previously unseen ligands. On the composition test set, the three-body KRR model achieved the lowest MAE of 2.75 kcal/mol, significantly outperforming the reference KRR model using standard-RACs with an MAE of 4.10 kcal/mol. However, all KRR models showed comparable generalization errors of roughly 4.8 to 5.0 kcal/mol on the ligand test set. For the three-body NN models, we observed a reduction in the MAE of about 0.7 kcal/mol compared to the reference model using standard-RACs on both the composition test set (3.41 kcal/mol vs. 4.16 kcal/mol) and the ligand test set (4.00 kcal/mol vs. 4.73 kcal/mol). Despite slightly higher MAEs of 3.61 kcal/mol on the composition test set and 4.14 kcal/mol on the ligand test set, the two-body NN model represents the best trade-off between accuracy, data efficiency, and ease of training, i.e., robustness to overfitting.

Finally, we investigated the model performance on the frontier orbital energies, i.e., HOMO energy, LUMO energy, and the HOMO–LUMO gap. On the composition test set, the MBE-based models achieved significantly lower MAEs than the standard-RACs reference models for all frontier orbital energy targets. For example, on the HOMO energies the two-body and three-body KRR model achieve MAEs of 0.28 eV and 0.23 eV, respectively, compared to a reference value of 0.58 eV for the standard-RACs KRR model. However, we also observed significant overfitting for the three-body NN resulting an increased composition test set MAE on HOMO energies that propagated to the gap predictions. On the ligand test set, all models exhibit generally high MAEs (0.8 eV to 1.2 eV) compared to the other data sets, and there was no significant difference between the featurization approaches. The only exception to this trend is a decrease of roughly 0.1 to 0.2 eV in the MAE for the gap predictions of the MBE-based models, which we attribute to improved error cancellation of the HOMO and LUMO predictions due to the more systematic featurization approach. The limited generalization behavior for frontier orbital energies is attributed to limitations in the handcrafted features used in both the standard-RACs and the MBE-based approach. One way to address this in future work would be by incorporating the presented MBE-based encoding of TMC stereochemistry into a deep learning approach such as graph convolutional neural networks. The observed trade-off between limited accuracy of the two-body models and overparametrization in the three-body model could be addressed by partial inclusion of three-body interactions, e.g., using only *cis*-type interactions but not *trans*-type interactions or only curvature parameters but not isomer-splitting parameters (Appendix A). Such extensions could be particularly useful in the extension of the presented framework to catalytic properties of TMCs and non-octahedral geometries.

APPENDIX

**A. Linear dependence of three-body interaction terms.**

Because the presented approach does not include explicit dependence of the interaction terms on geometric information, such as bond lengths or bond angles, the only geometric encoding in the truncated MBE is the distinction between *cis*-type and *trans*-type three-body interactions. Truncating equation 1 from the main text at the three-body order and separating the sum over three-body interactions into *cis*-type and *trans*-type yields,

$$Q = q_1(M) + \sum_{i=1}^{6} q_2(M, L_i) + \sum_{i,j \in cis} q_3^{cis}(M, L_i, L_j) + \sum_{i,j \in trans} q_3^{trans}(M, L_i, L_j). \quad (A1)$$

This summation over the MBE terms can be rewritten as a vector product of an interaction vector **q** that contains all possible interaction terms and a coefficient vector **n** that counts the number of interactions of each type for a given TMC.

For example, all ten complexes that can be built from the combination of a single metal center and two ligands $A$ and $B$ can be expressed as,

$$\begin{bmatrix} Q(M(L_A)_6) \\ Q(M(L_A)_5(L_B)_1) \\ Q(M(L_A)_4(L_B)_{2,cis}) \\ Q(M(L_A)_4(L_B)_{2,trans}) \\ Q(M(L_A)_3(L_B)_{3,fac}) \\ Q(M(L_A)_3(L_B)_{3,mer}) \\ Q(M(L_A)_2,cis(L_B)_4) \\ Q(M(L_A)_{2,trans}(L_B)_4) \\ Q(M(L_A)_1(L_B)_5) \\ Q(M(L_B)_6) \end{bmatrix} = \begin{bmatrix} 6 & 0 & 12 & 0 & 0 & 3 & 0 & 0 \\ 5 & 1 & 8 & 4 & 0 & 2 & 1 & 0 \\ 4 & 2 & 5 & 6 & 1 & 1 & 2 & 0 \\ 4 & 2 & 4 & 8 & 0 & 2 & 0 & 1 \\ 3 & 3 & 3 & 6 & 3 & 0 & 3 & 0 \\ 3 & 3 & 2 & 8 & 2 & 1 & 1 & 1 \\ 2 & 4 & 1 & 6 & 5 & 0 & 2 & 1 \\ 2 & 4 & 0 & 8 & 4 & 1 & 0 & 2 \\ 1 & 5 & 0 & 4 & 8 & 0 & 1 & 2 \\ 0 & 6 & 0 & 0 & 12 & 0 & 0 & 3 \end{bmatrix} \begin{bmatrix} q_2(M, L_A) \\ q_2(M, L_B) \\ q_3^{cis}(M, L_A, L_A) \\ q_3^{cis}(M, L_A, L_B) \\ q_3^{cis}(M, L_B, L_B) \\ q_3^{trans}(M, L_A, L_A) \\ q_3^{trans}(M, L_A, L_B) \\ q_3^{trans}(M, L_B, L_B) \end{bmatrix}, \quad (A2)$$

where the constant one-body term and other unused interaction terms are not listed explicitly for the sake of clarity. In this example, the (10 × 8) matrix of coefficient vectors has rank 4, indicating linear dependence between the coefficient vectors. This linear dependence is a consequence of the restriction to octahedral complexes that imposes additional constraints on the coefficient vector, such as the restriction that the number of two-body interactions sums to a total of six.

To illustrate how this finding generalizes beyond a single metal center and the combination of two ligands, we give examples of how these constraints allow one to calculate the coefficients from a "minimal" basis. Note, however, that this choice of basis is not unique. In octahedral complexes, the number of *cis*-type three-body interactions involving two ligands of the same type $X$ can be calculated from the number of two-body interactions for that ligand and the number of *cis*-type three-body interactions between $X$ and all other ligand types,

$$n_3^{cis}(M, L_X, L_X) = \frac{1}{2}\left(4n_2^{cis}(M, L_X) - \sum_{Y \neq X} n_3^{cis}(M, L_X, L_Y)\right). \tag{A3}$$

A simple interpretation of this relationship is that every ligand of type $X$ in a given complex, quantified by $n_2^{cis}(M, L_X)$, could potentially form *cis*-type interactions with all four nearest neighbor ligands. The actual number of *cis*-type interactions is reduced by one for every occurrence of an interaction between a $X$-type ligand and a ligand of any other type. Finally, the factor one half accounts for double counting of interactions. Similarly, $n_3^{trans}(M, L_X, L_X)$, the number of $q_3^{trans}(M, L_X, L_X)$ interactions, is given by,

$$n_3^{trans}(M, L_X, L_X) = \frac{1}{2}\left(n_2^{trans}(M, L_X) - \sum_{Y \neq X} n_3^{trans}(M, L_X, L_Y)\right). \tag{A4}$$

As a consequence, the four interaction terms $q_2(M, L_A)$, $q_2(M, L_B)$, $q_3^{cis}(M, L_A, L_B)$, and $q_3^{trans}(M, L_A, L_B)$ form a complete basis for the example from equation A2, i.e, restriction to these four terms results in a model of the same expressive power as the full model.

Another interesting finding from equation A2 is that the difference between expressions for the *cis* and *trans* isomers is exactly the same as the difference between the *fac* and *mer* isomer expressions,

$$
\begin{aligned}
\Delta q_3^{isomer}(M, L_A, L_B) &= Q\big(M(L_A)_4(L_B)_{2,cis}\big) - Q\big(M(L_A)_4(L_B)_{2,trans}\big) \\
&= Q\big(M(L_A)_3(L_B)_{3,fac}\big) - Q\big(M(L_A)_3(L_B)_{3,mer}\big) \\
&= Q\big(M(L_A)_{2,cis}(L_B)_4\big) - Q\big(M(L_A)_{2,trans}(L_B)_4\big) \\
&= q_3^{cis}(M, L_A, L_A) - 2q_3^{cis}(M, L_A, L_B) + q_3^{cis}(M, L_B, L_B) \\
&\quad - q_3^{trans}(M, L_A, L_A) + 2q_3^{trans}(M, L_A, L_B) - q_3^{trans}(M, L_B, L_B).
\end{aligned}
\tag{A5}
$$

In combination with the "curvature" of the interpolation curve, i.e., the difference between the expression for the 5+1 complexes and linear interpolation,

$$
\begin{aligned}
\Delta q_3^{5+1}(M, L_A, L_B) &= Q(M(L_A)_5(L_B)_1) - \left(\frac{5}{6}Q(M(L_A)_6) + \frac{1}{6}Q(M(L_B)_6)\right) \\
&= Q(M(L_A)_1(L_B)_5) - \left(\frac{1}{6}Q(M(L_A)_6) + \frac{5}{6}Q(M(L_B)_6)\right) \\
&= -2q_3^{cis}(M, L_A, L_A) + 4q_3^{cis}(M, L_A, L_B) - 2q_3^{cis}(M, L_B, L_B) \\
&\quad - \frac{1}{2}q_3^{trans}(M, L_A, L_A) + q_3^{trans}(M, L_A, L_B) - \frac{1}{2}q_3^{trans}(M, L_B, L_B).
\end{aligned}
\tag{A6}
$$

and the expressions for the two homoleptic complexes,

$$
\tilde{q}_2(M, L_A) = \frac{1}{6}Q(M(L_A)_6) = q_2(M, L_A) + 2q_3^{cis}(M, L_A, L_A) + \frac{1}{2}q_3^{trans}(M, L_A, L_A),
\tag{A7}
$$

and

$$\tilde{q}_2(M, L_B) = \frac{1}{6} Q(M(L_B)_6) = q_2(M, L_B) + 2q_3^{cis}(M, L_B, L_B) + \frac{1}{2} q_3^{trans}(M, L_B, L_B), \tag{A8}$$

this isomer-splitting term forms a basis that allows for easier interpretation of the effect of each interaction term.

**B. Derivation of MBE based kernels.**

Kernel ridge regression (KRR) is usually presented as an extension of linear ridge regression in which the input vector $\mathbf{x}_i$ is mapped to a $D$-dimensional feature space $\boldsymbol{\phi}_i = \boldsymbol{\phi}(\mathbf{x}_i)$, where $D$ is typically larger than the dimension of the original input space.[107,114] A linear model in this feature space is given by the expression,

$$Q(\mathbf{x}) = \sum_d^D w_{(d)} \phi_{(d)}(\mathbf{x}) = \mathbf{w}^\top \boldsymbol{\phi}(\mathbf{x}), \tag{B1}$$

where $\mathbf{w}^\top = [w_{(1)}, ..., w_{(D)}]$ is the vector of weights for the linear regression. This weight vector is determined by minimizing the ridge regression cost function,

$$\mathcal{L} = \frac{1}{2} \sum_n^N (Q_n - \mathbf{w}^\top \boldsymbol{\phi}(\mathbf{x}_n))^2 + \frac{1}{2} \lambda \|\mathbf{w}\|_2^2, \tag{B2}$$

where the pairs $(\mathbf{x}_1, Q_1), ..., (\mathbf{x}_N, Q_N)$ are the training data and $\lambda$ is a hyperparameter tuning the strength of the $L_2$ regularization term. Setting the derivative of the cost function with respect to the weight vector to zero yields,

$$\sum_n^N (Q_n - \mathbf{w}^\top \boldsymbol{\phi}_n) \boldsymbol{\phi}_n = \lambda \mathbf{w}, \tag{B3}$$

which can be rearranged into a closed form expression for the weight vector,

$$\mathbf{w} = \left(\lambda \mathbf{I}_D + \sum_n \boldsymbol{\phi}_n \boldsymbol{\phi}_n^\top\right)^{-1} \left(\sum_n Q_n \boldsymbol{\phi}_n\right), \tag{B4}$$

where $\mathbf{I}_D$ is an identity matrix of dimension $(D \times D)$. Using the definitions $\Phi = [\boldsymbol{\phi}_1, ..., \boldsymbol{\phi}_N]$ and $\mathbf{Q} = [Q_1, ..., Q_N]^\top$, the sums can be rewritten in terms of vector/matrix multiplications,

$$\mathbf{w} = (\lambda \mathbf{I}_D + \Phi \Phi^\top)^{-1} \Phi \mathbf{Q}. \tag{B5}$$

The expression for the weights can be further rearranged by applying the following matrix identity, often referred to as the push-through identity,[115-117]

$$(A + P^\top R^{-1} P)^{-1} P^\top R^{-1} = A^{-1} P^\top (R + P A^{-1} P^\top)^{-1}, \tag{B6}$$

where setting $A = \lambda \mathbf{I}_D$, $P^\top = \Phi$, and $R = \mathbf{I}_N$ yields,

$$\mathbf{w} = \Phi(\lambda \mathbf{I}_N + \Phi^\top \Phi)^{-1} \mathbf{Q}. \tag{B7}$$

This allows one to perform the costly matrix inversion either in the $(D \times D)$-dimensional feature space or the $(N \times N)$-dimensional space of training examples. This switch from outer product in feature space to inner product also gives rise to the so-called kernel trick, where the inner product in feature space is replaced with a kernel function $\boldsymbol{\phi}(\mathbf{x})^\top \boldsymbol{\phi}(\mathbf{y}) = k(\mathbf{x}, \mathbf{y})$. Plugging the expression for the weights in training example space (eq. B7) back into the definition of the linear model (eq. B1) yields,

$$Q(\mathbf{x}) = \mathbf{Q}^\top (\Phi^\top \Phi + \lambda \mathbf{I}_N)^{-1} \Phi^\top \boldsymbol{\phi}(\mathbf{x}). \tag{B8}$$

Finally, applying the kernel trick gives,

$$Q(\mathbf{x}) = \mathbf{Q}^\top (K + \lambda \mathbf{I}_N)^{-1} \mathbf{k}(\mathbf{x}) \tag{B9}$$

where **k(x)** is a vector obtained by evaluating the kernel function for **x** and all training examples $\mathbf{x}_n$,

$$\mathbf{k}(\mathbf{x}) = \Phi^\top \boldsymbol{\phi}(\mathbf{x}) = \begin{bmatrix} k(\mathbf{x}_1, \mathbf{x}) \\ \vdots \\ k(\mathbf{x}_N, \mathbf{x}) \end{bmatrix}, \tag{B10}$$

and $K$ is the matrix of kernel values of pairwise combinations of training examples,

$$K = \Phi^\top \Phi = \begin{bmatrix} k(\mathbf{x}_1, \mathbf{x}_1) & \cdots & k(\mathbf{x}_1, \mathbf{x}_N) \\ \vdots & \ddots & \vdots \\ k(\mathbf{x}_N, \mathbf{x}_1) & \cdots & k(\mathbf{x}_N, \mathbf{x}_N) \end{bmatrix}. \tag{B11}$$

To derive a KRR expression for the MBE based models, we start with a two-body model in weight-space view given by,

$$Q^{\text{KRR}}(\mathbf{M}, \mathbf{L}_1, \dots, \mathbf{L}_6) = q_1(\mathbf{M}) + \sum_{i=1}^{6} q_2(\mathbf{M}, \mathbf{L}_i) = \mathbf{w}_1^\top \boldsymbol{\phi}_1(\mathbf{M}) + \sum_{i=1}^{6} \mathbf{w}_2^\top \boldsymbol{\phi}_2(\mathbf{M}, \mathbf{L}_i) \tag{B12}$$

where $\mathbf{w}_1$ and $\mathbf{w}_2$ are weight vectors and $\boldsymbol{\phi}_1$ and $\boldsymbol{\phi}_2$ are transformations from the respective vectorial input space to the feature space. Importantly, this expression can be rearranged into the standard KRR weight space view by stacking the weight and transformation vectors,

$$Q^{\text{KRR}}(\mathbf{M}, \mathbf{L}_1, \dots, \mathbf{L}_6) = \begin{bmatrix} \mathbf{w}_1 \\ \mathbf{w}_2 \end{bmatrix}^\top \begin{bmatrix} \boldsymbol{\phi}_1(\mathbf{M}) \\ \sum_{i=1}^{6} \boldsymbol{\phi}_2(\mathbf{M}, \mathbf{L}_i) \end{bmatrix} = \mathbf{w}^\top \boldsymbol{\phi}(\mathbf{x}), \tag{B13}$$

and, therefore, results in exactly the same prediction expression (eq. B8). However, special care needs to be taken when applying the kernel trick. Since the kernel function $k(\mathbf{x}, \mathbf{y})$ is used to replace the inner product $\boldsymbol{\phi}(\mathbf{x})^\top \boldsymbol{\phi}(\mathbf{y})$, all properties of this inner product must be conserved. In our specific case, the important properties are the separation into the individual MBE contributions and the summation over two-body terms,

$$\begin{aligned}\boldsymbol{\phi}(\mathbf{x})^\top \boldsymbol{\phi}(\mathbf{y}) &= \begin{bmatrix} \boldsymbol{\phi}_1(\mathbf{M}_x) \\ \sum_{i=1}^{6} \boldsymbol{\phi}_2(\mathbf{M}_x, \mathbf{L}_{x,i}) \end{bmatrix}^\top \begin{bmatrix} \boldsymbol{\phi}_1(\mathbf{M}_y) \\ \sum_{i=1}^{6} \boldsymbol{\phi}_2(\mathbf{M}_y, \mathbf{L}_{y,j}) \end{bmatrix} \\ &= \boldsymbol{\phi}_1(\mathbf{M}_x)^\top \boldsymbol{\phi}_1(\mathbf{M}_y) + \sum_{i=1}^{6}\sum_{j=1}^{6} \boldsymbol{\phi}_2(\mathbf{M}_x, \mathbf{L}_{x,i})^\top \boldsymbol{\phi}_2(\mathbf{M}_y, \mathbf{L}_{y,j}). \end{aligned} \quad (B14)$$

Applying the kernel trick separately to each of these inner products of transformation vectors yields,

$$k(\mathbf{x},\mathbf{y}) = k_1(\mathbf{M}_x, \mathbf{M}_y) + \sum_{i=1}^{6}\sum_{j=1}^{6} k_2\big([\mathbf{M}_x, \mathbf{L}_{x,i}], [\mathbf{M}_y, \mathbf{L}_{y,j}]\big). \quad (B15)$$

The same arguments can be used to derive a KRR expression for the three-body model by adding the *cis*-type and *trans*-type interaction terms.

**SUPPLEMENTARY MATERIAL**.

See the supplementary material for details on the geometry metrics and thresholds used in this work; composition of the curated data set; composition of the ligand generalization test set; description of the validation set; scatter plot matrix of the four frontier orbital energies; schematic of the MBE-based neural network; optimized NN hyperparameters for the SSE; optimized NN hyperparameters for the frontier orbital energies; optimized KRR hyperparameters for the SSE; optimized KRR hyperparameters for the frontier orbital energies; SSE MAEs for metal/oxidation state subsets of the validation set; histograms of the DFT spin-splitting energies; $R^2$ scores of the ML predictions of the SSE; SSE $R^2$ scores for metal/oxidation state subsets of the validation set; parity plot of SSE predictions vs. target on the training set; SSE MAEs for metal/oxidation state subsets of the composition test set; SSE $R^2$ scores for metal/oxidation state subsets of the composition test set; PCA of the two RACs feature spaces on the training set; interpolation plot of the KRR SSE predictions on the composition test set; interpolation plot of the NN SSE predictions on the composition test set; SSE MAEs for metal/oxidation state subsets of the ligand test set; SSE MAEs for ligand subsets of the ligand test set; average negative log-likelihood values of the SSE predictions; histograms of the DFT HOMO energies; total charge of all six ligands of the complexes in the training set; $R^2$ scores for the ML predictions of the HOMO energies; composition dependence of the HOMO MAEs on the composition test set; interpolation plot of DFT target data for the LS HOMO energies; interpolation plot of KRR HOMO predictions on the composition test set; HOMO MAEs for metal/oxidation state subsets of the ligand test set: HOMO MAEs for ligand subsets of the ligand test set; HOMO/LUMO prediction errors for ligand subsets of the ligand test set; HOMO STD of absolute errors for ligand subsets of the ligand test set; analysis of the ML predictions for LUMO energies; LUMO MAEs on all four data sets; composition dependence of LUMO MAEs on the composition test set; histogram of the DFT LUMO energies; LUMO DFT target data for the composition test set; histogram of the DFT gap energies; parity plot of LS vs. HS DFT gap energies on the training set; correlation metrics of the LS vs. HS DFT gap energies; number of negative gap energy predictions on all four data sets; $R^2$ scores for the ML predictions of the gap energies; MAEs of gap energy models trained directly on the DFT gap data. (PDF)

AUTHOR INFORMATION

**Corresponding Author**

*email:hjkulik@mit.edu

**Notes**

The authors declare no competing financial interest.

ACKNOWLEDGMENT

Initial support for this work, including data curation and initial model testing, was provided by the Office of Naval Research under grant number N00014-20-1-2150. Subsequent support for final model training and development was provided by the United States Department of Energy under grant number DE-NA0003965. D.B.K.C. was partially supported by a National Science Foundation Graduate Research Fellowship Program under grant #1745302. H.J.K. acknowledges a Sloan Foundation Fellowship in Chemistry and a Simon Family Faculty Research Innovation Fund grant. The authors thank Adam Steeves, Aditya Nandy, and Vyshnavi Vennelakanti for helpful discussions.

DATA AVAILABILITY

The data that support the findings of this study are openly available in the Supporting Information associated with this article and at the following URL and DOI: https://github.com/hjkgrp/many_body_ml and https://doi.org/10.5281/zenodo.13331586.

**For Table of Contents Use Only**

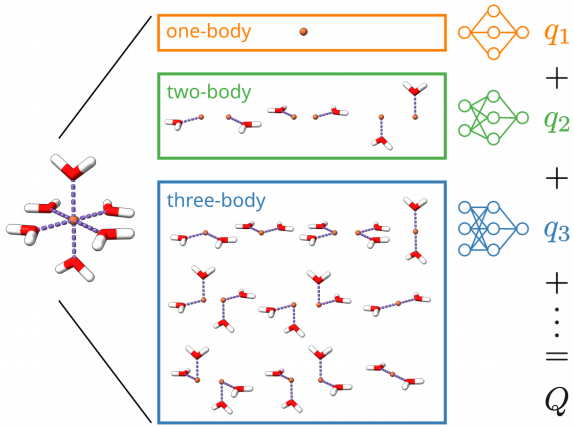

# Supplementary Material for

## *Many-body Expansion Based Machine Learning Models for Octahedral Transition Metal Complexes*

Ralf Meyer[1], Daniel B. K. Chu[1], and Heather J. Kulik[1,2,*]

[1]Department of Chemical Engineering, Massachusetts Institute of Technology, Cambridge, MA 02139, USA

[2]Department of Chemistry, Massachusetts Institute of Technology, Cambridge, MA 02139, USA

*Corresponding author email: hjkulik@mit.edu

**Contents**

Table S1. The geometry metrics and thresholds used in this work are unmodified from our prior work[1] and include: full connectivity of the complex, a coordination number of 6 (atoms are considered bonded if their distance is below a threshold defined by the sum of covalent radii), thresholds on mean and maximum deviations of the connecting-atom–metal–connecting-atom angles, thresholds on the maximum deviation of metal–ligand bond lengths for all ligands and for just the equatorial ligands, and thresholds on the mean and maximum for angles between the metal center and first two atoms in linear ligands.

| Category | Metric | Threshold |
|---|---|---|
| Connectivity | Bond between atoms a and b | $d(ab) < 1.37 * (r_{cov}(a) + r_{cov}(b))$ |
| | Fully connected | = True |
| | Coordination number | = 6 |
| Shape of first coordination sphere | $mean(\Delta\theta(C_i\text{-}M\text{-}C_j))$ | < 12° |
| | $max(\Delta\theta(C_i\text{-}M\text{-}C_j))$ | < 22.5° |
| | $max(\Delta d)$ | < 1 Å |
| | $max(\Delta d_{eq})$ | < 0.25 Å |
| Angles with linear ligands | $mean(\Delta\theta(M\text{-}A\text{-}B))$ | < 20° |
| | $max(\Delta\theta(M\text{-}A\text{-}B))$ | < 28° |

Table S2. Composition of the curated data set after applying each of the three checks for negative majority-spin highest occupied molecular orbital energy, deviation from octahedral geometry, and deviation of the spin expectation value. Each row corresponds to the number of data points that remain after removal due to the criterion listed on the left.

| | Cr(III) | Cr(II) | Mn(III) | Mn(II) | Fe(III) | Fe(II) | Co(III) | Co(II) | all |
|---|---|---|---|---|---|---|---|---|---|
| initial data set | 351 | 245 | 272 | 341 | 309 | 321 | 220 | 291 | 2350 |
| negative majority-spin HOMO | 273 | 211 | 232 | 298 | 261 | 286 | 172 | 268 | 2001 |
| geometry check | 271 | 203 | 217 | 286 | 257 | 280 | 165 | 263 | 1942 |
| spin check (final) | 264 | 179 | 162 | 265 | 248 | 275 | 163 | 250 | 1806 |

Table S3. Composition of the ligand generalization test set after applying each of the three checks for negative majority-spin highest occupied molecular orbital energy, deviation from octahedral geometry, and deviation of the spin expectation value. Each row corresponds to the number of data points that remain after removal due to the criterion listed on the left.

| | Cr(III) | Cr(II) | Mn(III) | Mn(II) | Fe(III) | Fe(II) | Co(III) | Co(II) | all |
|---|---|---|---|---|---|---|---|---|---|
| initial data set | 21 | 21 | 21 | 21 | 21 | 21 | 21 | 21 | 168 |
| negative majority-spin HOMO | 21 | 21 | 21 | 21 | 21 | 21 | 21 | 21 | 168 |
| geometry check | 20 | 19 | 19 | 20 | 17 | 21 | 14 | 21 | 151 |
| spin check (final) | 18 | 18 | 9 | 18 | 14 | 21 | 8 | 21 | 127 |

**Text S1.** Description of the validation set.
The validation set was constructed by first selecting 11 ligands based on the atomic element of their connecting atoms and placing every complex containing at least one of these ligands into the validation set. In an effort to represent their relative abundance in the overall data set, this set of 11 ligands consists of three oxygen- and three nitrogen-coordinating ligands, two sulfur- and two phosphor-coordinating ligands, and one carbon-coordinating ligand with the SMILES strings listed below:

- oxygen-coordinating: [O]=[CH2], [O]=[PH], [OH]-[NH]-[NH]-[OH]
- nitrogen-coordinating: [NH2]-[S]-[S]-[NH2], [NH]=[CH]-[CH]=[NH], [NH]=[NH],
- sulfur-coordinating: [SH]-[CH3], [SH]-[CH]=[CH]-[SH],
- phosphor-coordinating: [PH]=[CH2], [PH2]-[CH2]-[CH2]-[PH2],
- carbon-coordinating: [C+]=[CH2-]

This resulted in an initial validation set size of 156 complexes. The remaining 206 complexes to achieve a rough 80/20 split were drawn randomly while ensuring that homoleptic complexes of the three ligands used to generate the composition test set remain in the training set.

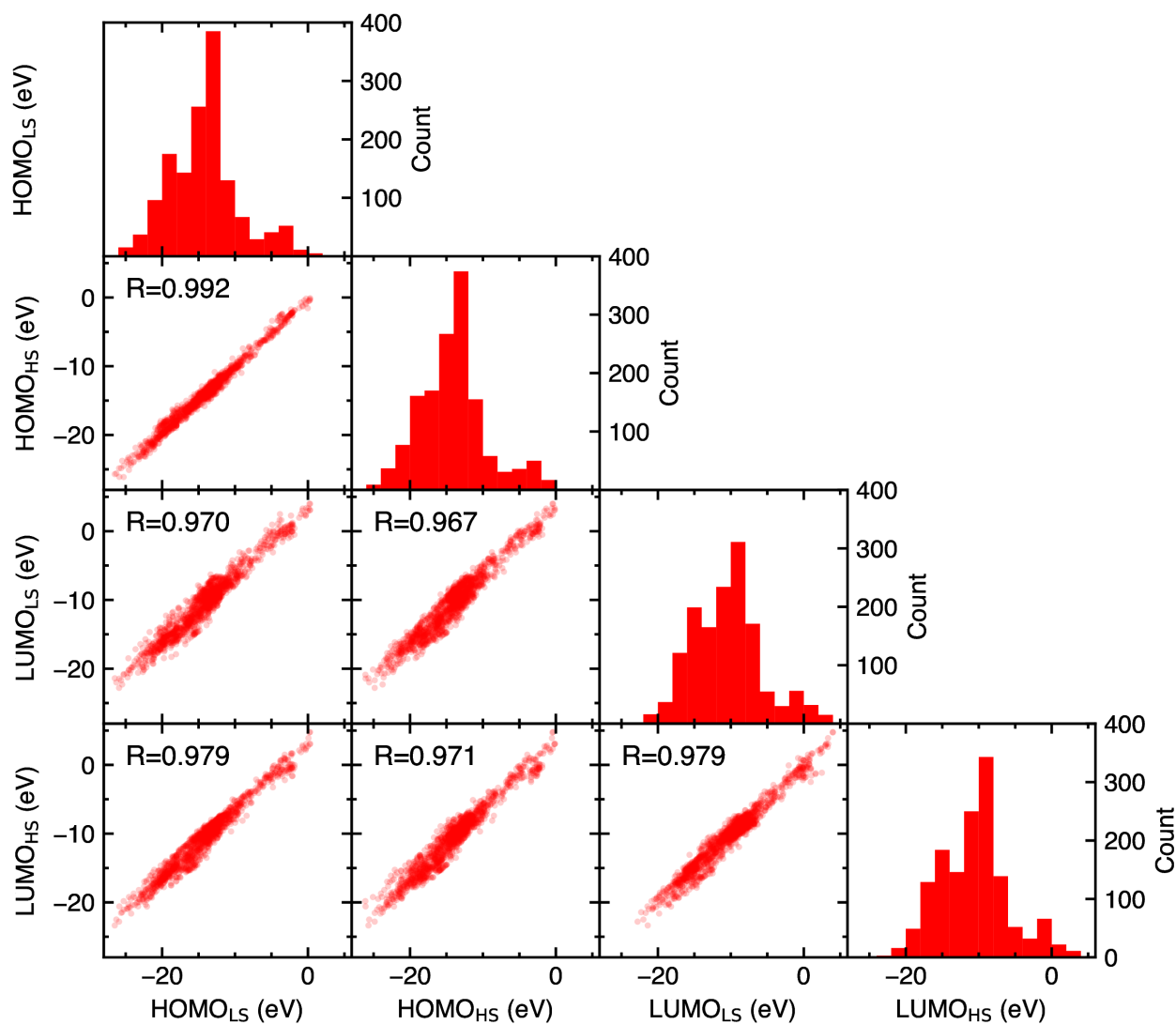

**Figure S1.** Scatter plot matrix of the four frontier orbital energies for the 1,444 training set complexes. This plot consists of histograms for each of the frontier orbital energies with a bin width of 2 eV (shown on the diagonal) and scatter plots and Pearson correlation coefficients for each pairwise combination of orbital energies (shown on the off-diagonal).

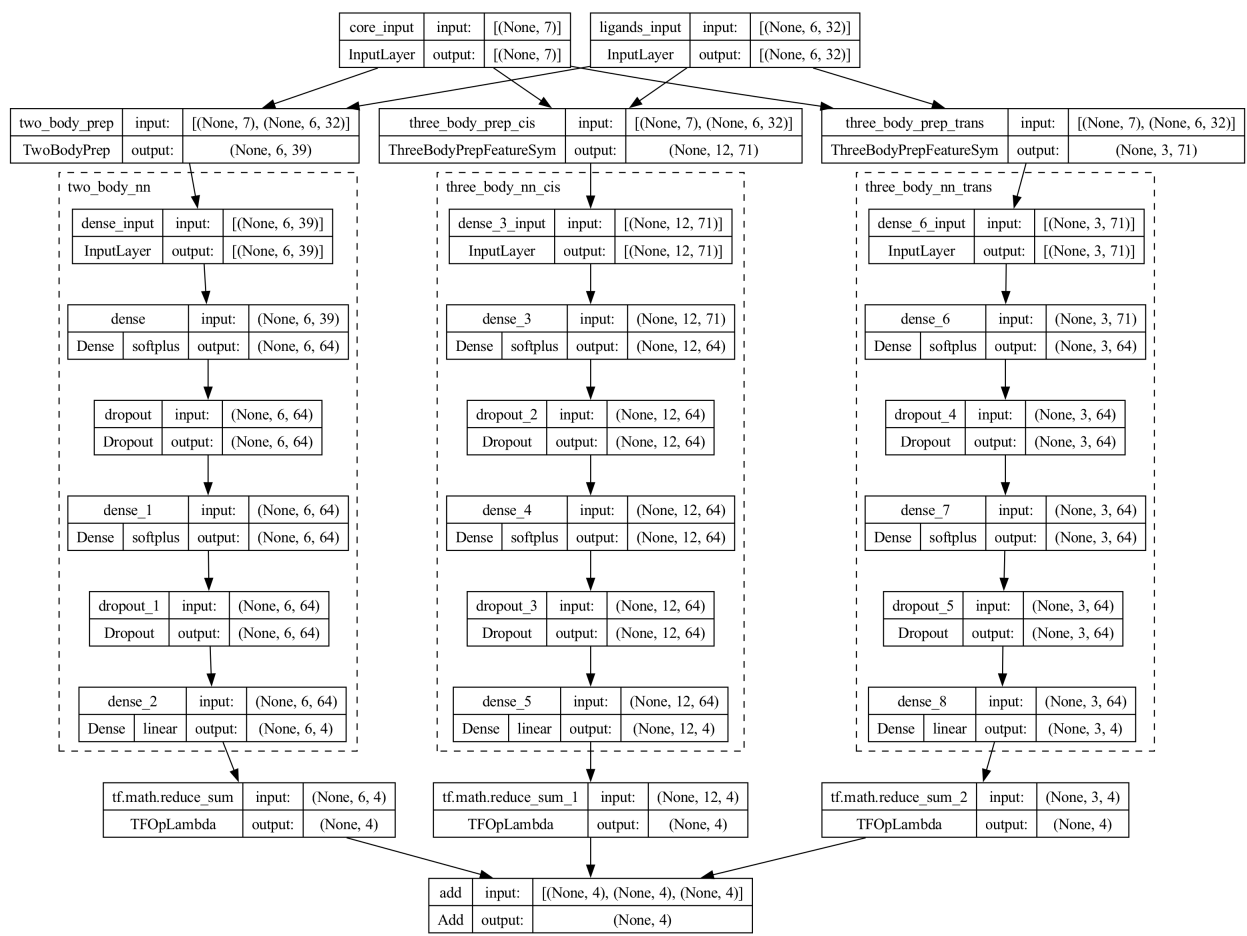

**Figure S2.** Schematic of the MBE-based neural network architecture as used in the multi-task learning of the four frontier orbital energies. The feed-forward architecture, plotted from top to bottom, consists of two input layers for the metal core and ligand features, three layers to preprocess the features by concatenating and/or symmetrizing, followed by three multilayer perceptrons used to model the interaction terms, layers that sum over contributions of the same type, and, finally, a layer that sums over the three different interaction terms. This architecture does not include an explicit one-body term and instead relies on the higher-order MBE terms to capture the metal core dependence.

**Table S4.** Optimized hyperparameters of the NN models for the spin-splitting energy.

| Model | Batch size | Dropout probability | L2-regularization | NN architecture |
|---|---|---|---|---|
| NN standard-RACs | 64 | 0.0 | $10^{-4}$ | (256, 256) |
| NN two-body | 64 | 0.1 | $10^{-5}$ | (256, 256) |
| NN three-body | 128 | 0.1 | $10^{-5}$ | (256, 256) |

**Table S5.** Optimized hyperparameters of the NN models for the frontier orbital energies.

| Model | Batch size | Dropout probability | L2-regularization | NN architecture |
|---|---|---|---|---|
| NN standard-RACs | 64 | 0.1 | $10^{-5}$ | (256, 256) |
| NN two-body | 64 | 0.0 | $10^{-5}$ | (128, 128) |
| NN three-body | 64 | 0.1 | $10^{-4}$ | (256, 256) |

**Table S6.** Optimized hyperparameters of the KRR models for the spin-splitting energy. For the three-body model the parameters of the two-body kernel are set to the values obtained in the hyperparameter-optimization of the two-body model.

| Model | L2-regularization | Kernel parameters |
|---|---|---|
| KRR standard-RACs | 0.10 | $c_{\text{RACs}} = 104.4, k_{\text{RACs}} = k^{\text{Matérn}}(l = 8.3)$ |
| KRR two-body | 3.91 | $c_2 = 215.8, k_2 = k^{\text{Matérn}}(l = 2.2)$ |
| KRR three-body | 0.29 | $c_{3,cis} = 26.7, k_{3,cis} = k^{\text{Matérn}}(l = 1.1),$ $c_{3,trans} = 1.8, k_{3,trans} = k^{\text{Matérn}}(l = 3.7)$ |

**Table S7.** Optimized hyperparameters of the KRR models for the frontier orbital energies. For the three-body model the parameters of the two-body kernel are set to the values obtained in the hyperparameter-optimization of the two-body model.

| Model | L2-regularization | Kernel parameters |
|---|---|---|
| KRR standard-RACs | 0.18 | $c_{\text{RACs}} = 119.6, k_{\text{RACs}} = k^{\text{RBF}}(l = 3.7)$ |
| KRR two-body | 0.10 | $c_2 = 375.1, k_2 = k^{\text{RBF}}(l = 4.3)$ |
| KRR three-body | 0.10 | $c_{3,cis} = 0.31, k_{3,cis} = k^{\text{Matérn}}(l = 0.21),$ $c_{3,trans} = 0.01, k_{3,trans} = k^{\text{RBF}}(l = 0.01)$ |

**Table S8.** Mean absolute error of the model predictions of the spin-splitting energies on subsets of the validation set for each of the eight metal/oxidation state combinations in kcal/mol.

| Model | Cr(III) | Cr(II) | Mn(III) | Mn(II) | Fe(III) | Fe(II) | Co(III) | Co(II) |
|---|---|---|---|---|---|---|---|---|
| KRR standard-RACs | 3.29 | 3.59 | 4.55 | 3.07 | 3.22 | 3.83 | 5.04 | 3.74 |
| KRR two-body | 2.13 | 4.96 | 3.90 | 3.70 | 3.98 | 5.03 | 4.31 | 3.85 |
| KRR three-body | 2.07 | 3.65 | 2.80 | 3.00 | 3.47 | 3.90 | 3.83 | 3.54 |
| NN standard-RACs | 2.35 | 4.05 | 4.04 | 3.35 | 3.51 | 3.43 | 4.60 | 3.55 |
| NN two-body | 2.17 | 4.31 | 4.08 | 3.49 | 3.76 | 4.28 | 4.72 | 3.60 |
| NN three-body | 2.12 | 4.27 | 4.32 | 3.07 | 3.46 | 3.70 | 4.56 | 3.33 |

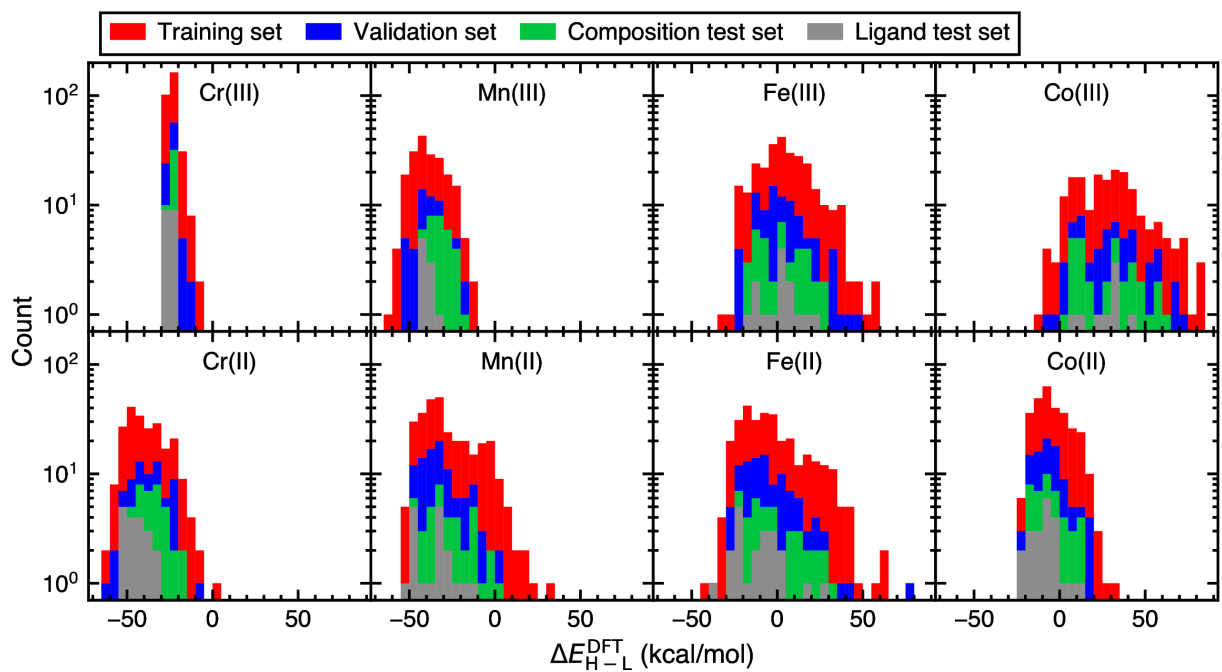

**Figure S3.** Stacked histogram of the DFT spin-splitting energies for all four data sets and eight metal/oxidation state combinations with a bin width of 5 kcal/mol grouped by set origin as indicated in inset legend.

**Table S9**. Coefficients of determination $R^2$ of the model predictions of the spin-splitting energies on all four data sets.

| Model | Training set | Validation set | Composition test set | Ligand test set |
|---|---|---|---|---|
| KRR standard-RACs | 0.998 | 0.959 | 0.947 | 0.921 |
| KRR two-body | 0.979 | 0.952 | 0.967 | 0.903 |
| KRR three-body | 0.996 | 0.962 | 0.976 | 0.907 |
| NN standard-RACs | 0.985 | 0.964 | 0.954 | 0.923 |
| NN two-body | 0.972 | 0.958 | 0.964 | 0.934 |
| NN three-body | 0.979 | 0.963 | 0.961 | 0.943 |

**Table S10.** Coefficients of determination $R^2$ of the model predictions of the spin-splitting energies on subsets of the validation set for each of the eight metal/oxidation state combinations.

| Model | Cr(III) | Cr(II) | Mn(III) | Mn(II) | Fe(III) | Fe(II) | Co(III) | Co(II) |
|---|---|---|---|---|---|---|---|---|
| KRR standard-RACs | -0.483 | 0.857 | 0.549 | 0.865 | 0.927 | 0.916 | 0.890 | 0.746 |
| KRR two-body | 0.199 | 0.754 | 0.661 | 0.798 | 0.866 | 0.869 | 0.929 | 0.747 |
| KRR three-body | 0.276 | 0.835 | 0.777 | 0.848 | 0.894 | 0.901 | 0.934 | 0.766 |
| NN standard-RACs | 0.241 | 0.828 | 0.680 | 0.846 | 0.905 | 0.941 | 0.911 | 0.765 |
| NN two-body | 0.274 | 0.770 | 0.667 | 0.809 | 0.888 | 0.911 | 0.913 | 0.775 |
| NN three-body | 0.359 | 0.796 | 0.623 | 0.856 | 0.902 | 0.930 | 0.920 | 0.786 |

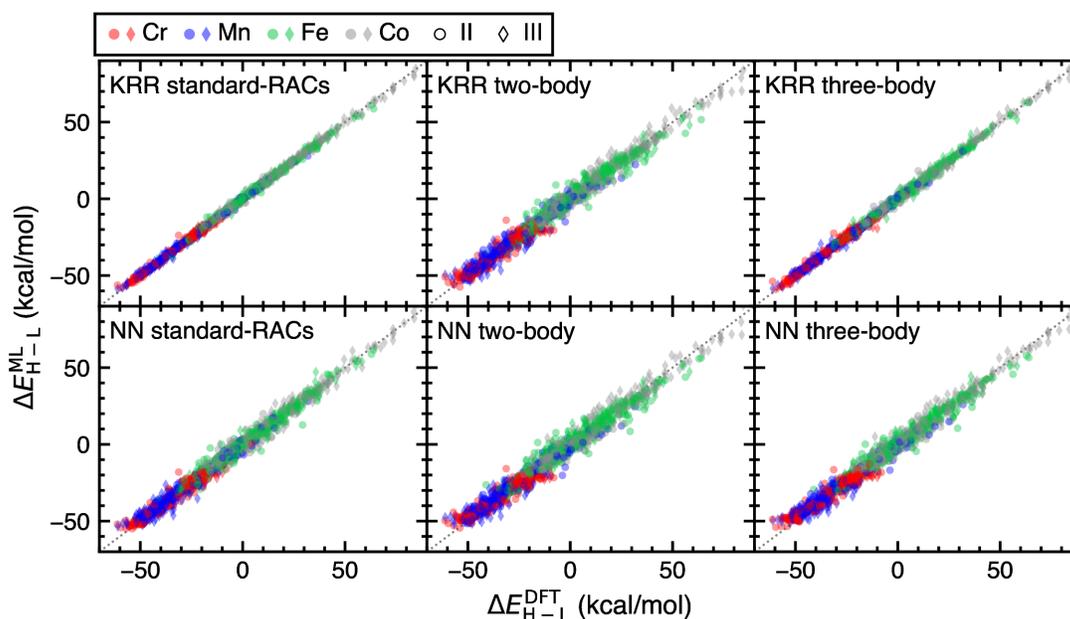

**Figure S4.** Calculated versus predicted spin-splitting energies of the training set complexes for the two ML methods (columns) and the three featurization approaches (rows).

**Table S11.** Mean absolute error of the model predictions of the spin-splitting energies on subsets of the composition test set for each of the eight metal/oxidation state combinations in kcal/mol.

| Model | Cr(III) | Cr(II) | Mn(III) | Mn(II) | Fe(III) | Fe(II) | Co(III) | Co(II) |
|---|---|---|---|---|---|---|---|---|
| KRR standard-RACs | 3.47 | 2.21 | 4.70 | 2.11 | 2.87 | 5.47 | 7.67 | 4.30 |
| KRR two-body | 1.36 | 3.08 | 5.78 | 1.48 | 1.48 | 3.56 | 5.95 | 2.24 |
| KRR three-body | 1.41 | 2.66 | 5.66 | 1.76 | 2.98 | 1.57 | 3.63 | 2.29 |
| NN standard-RACs | 2.57 | 3.76 | 7.30 | 2.85 | 2.87 | 4.50 | 4.72 | 4.70 |
| NN two-body | 2.36 | 2.82 | 6.42 | 3.48 | 2.28 | 3.07 | 4.45 | 3.99 |
| NN three-body | 2.17 | 4.39 | 8.45 | 1.99 | 1.98 | 1.72 | 3.25 | 3.35 |

**Table S12.** Coefficients of determination $R^2$ of the model predictions of the spin-splitting energies on subsets of the composition test set for each of the eight metal/oxidation state combinations.

| Model | Cr(III) | Cr(II) | Mn(III) | Mn(II) | Fe(III) | Fe(II) | Co(III) | Co(II) |
|---|---|---|---|---|---|---|---|---|
| KRR standard-RACs | -8.51 | 0.85 | 0.16 | 0.96 | 0.93 | 0.83 | 0.77 | 0.71 |
| KRR two-body | -0.66 | 0.77 | -0.16 | 0.98 | 0.98 | 0.93 | 0.86 | 0.92 |
| KRR three-body | -0.66 | 0.79 | -0.15 | 0.97 | 0.95 | 0.99 | 0.94 | 0.91 |
| NN standard-RACs | -3.48 | 0.66 | -0.92 | 0.94 | 0.94 | 0.92 | 0.91 | 0.68 |
| NN two-body | -2.94 | 0.79 | -0.41 | 0.92 | 0.96 | 0.95 | 0.92 | 0.78 |
| NN three-body | -2.45 | 0.57 | -1.56 | 0.97 | 0.97 | 0.98 | 0.96 | 0.83 |

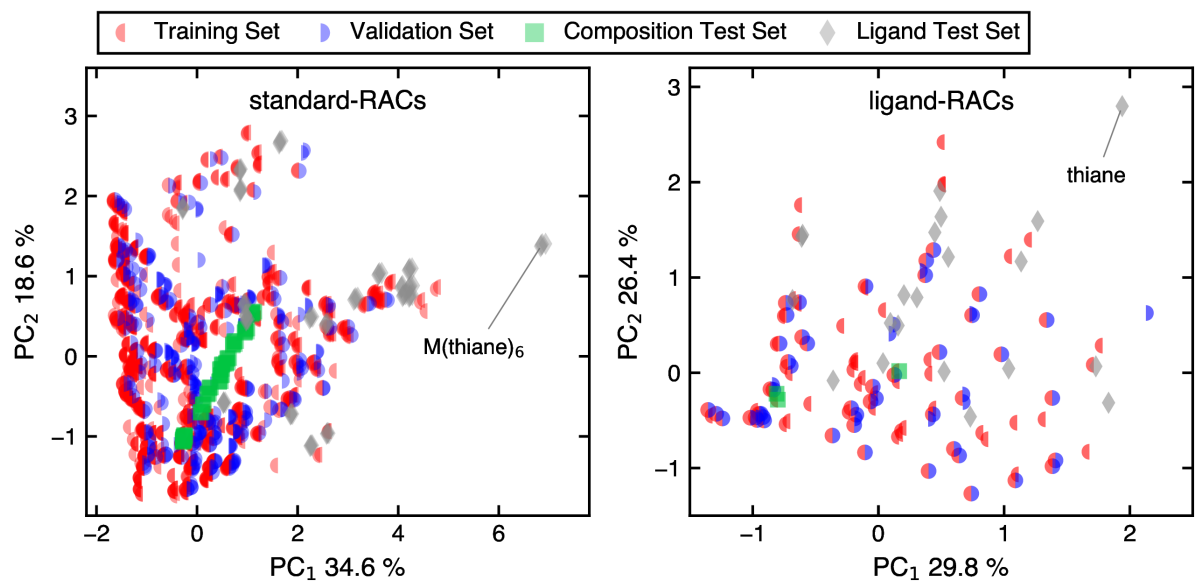

**Figure S5.** Two-dimensional principal component analysis of the two RACs feature spaces on the training set. Explained variance ratios for the two principal components are given in the axis labels. The feature vectors of the three other data sets are transformed into the same vector space.

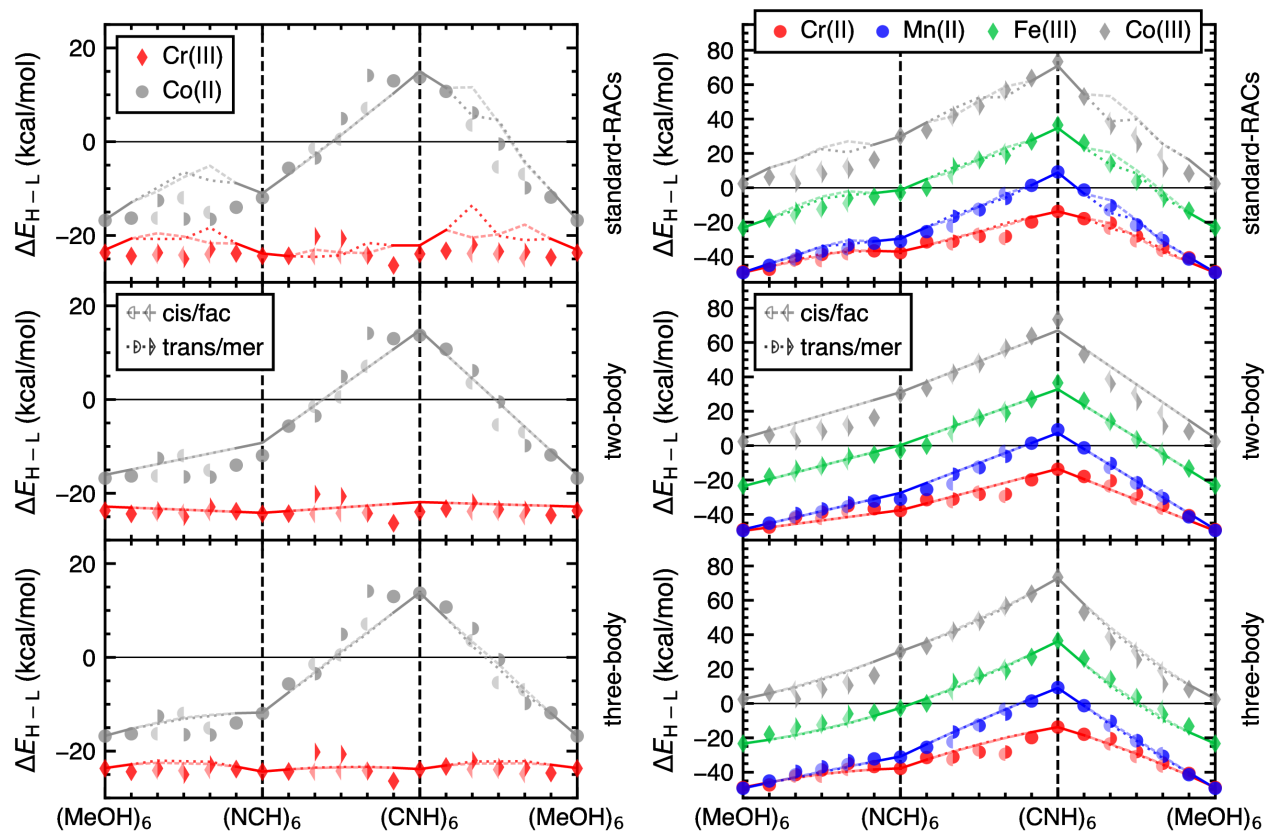

**Figure S6.** Plot of the spin-splitting energy interpolation curves for binary complexes of three ligands, methanol (MeOH), hydrogen cyanide (NCH), and hydrogen isocyanide (CNH), showing the DFT reference values (scatter) and the corresponding ML predictions (lines) for KRR models using different featurization approaches (rows). For compositions with multiple structural isomers, values are plotted using half markers for the DFT reference and dashed/dotted lines for the ML predictions.

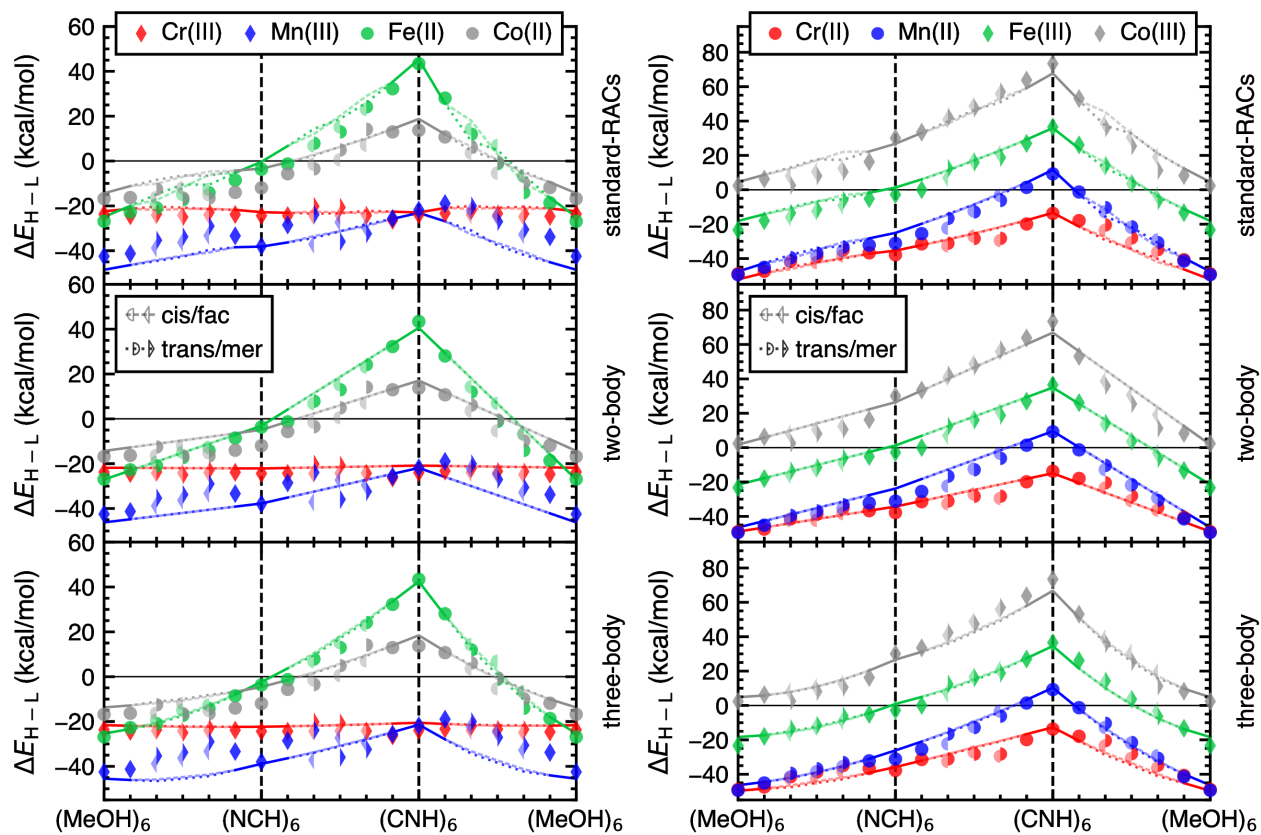

**Figure S7.** Plot of the spin-splitting energy interpolation curves for binary complexes of three ligands, methanol (MeOH), hydrogen cyanide (NCH), and hydrogen isocyanide (CNH), showing the DFT reference values (scatter) and the corresponding ML predictions (lines) for NN models using different featurization approaches (rows). For compositions with multiple structural isomers, values are plotted using half markers for the DFT reference and dashed/dotted lines for the ML predictions.

**Table S13.** Mean absolute error of the model predictions of the spin-splitting energies on subsets of the ligand test set for each of the eight metal/oxidation state combinations in kcal/mol.

| Model | Cr(III) | Cr(II) | Mn(III) | Mn(II) | Fe(III) | Fe(II) | Co(III) | Co(II) |
|---|---|---|---|---|---|---|---|---|
| KRR standard-RACs | 5.21 | 4.29 | 5.85 | 4.83 | 4.10 | 4.74 | 7.72 | 4.07 |
| KRR two-body | 3.74 | 3.35 | 4.13 | 6.68 | 5.15 | 6.95 | 8.62 | 3.31 |
| KRR three-body | 4.41 | 3.27 | 3.54 | 6.55 | 4.67 | 6.61 | 7.72 | 3.42 |
| NN standard-RACs | 2.54 | 3.46 | 5.37 | 5.00 | 6.54 | 4.98 | 9.32 | 3.97 |
| NN two-body | 1.18 | 2.76 | 5.15 | 5.32 | 5.21 | 4.91 | 8.32 | 3.33 |
| NN three-body | 1.26 | 3.06 | 3.79 | 5.17 | 5.17 | 4.88 | 7.83 | 3.14 |

**Table S14.** Mean absolute error of the model predictions of the spin-splitting energies on subsets of the ligand test set for each of the 21 ligands in kcal/mol. The largest error for each model is indicated in bold.

| Ligand | KRR | | | NN | | |
|---|---|---|---|---|---|---|
| | standard-RACs | two-body | three-body | standard-RACs | two-body | three-body |
| 4H-pyran | 4.40 | 2.50 | 3.33 | 4.14 | 2.46 | 4.08 |
| ethenediol | 7.78 | 1.77 | 2.97 | 3.83 | 3.02 | 3.85 |
| bifuran | 4.18 | 5.82 | 5.75 | 3.79 | 4.71 | 4.57 |
| pyridine-N-oxide | 3.27 | 4.01 | 3.84 | 3.98 | 3.19 | 3.59 |
| acrylamide | 7.02 | 6.83 | 6.83 | 2.56 | 4.23 | 3.28 |
| dimethylformamide | 2.79 | 4.37 | 3.89 | 2.72 | 1.76 | 1.85 |
| thiophene | 4.58 | 5.44 | 5.61 | 2.92 | 3.07 | 2.68 |
| thiane | 6.16 | 9.66 | 10.20 | **10.26** | **11.21** | **10.29** |
| 4H-thiopyran | 4.11 | 4.26 | 4.37 | 6.77 | 3.81 | 4.23 |
| oxazoline | 3.84 | 4.85 | 4.92 | 5.44 | 5.17 | 4.38 |
| thioazole | 4.12 | 4.60 | 4.80 | 5.36 | 3.22 | 3.62 |
| NHCHOH | 4.14 | 2.23 | 1.66 | 3.81 | 3.17 | 2.91 |
| PHCHOH | **9.08** | **12.74** | **11.09** | 8.55 | 7.62 | 6.72 |
| tetrazane | 2.82 | 3.58 | 3.78 | 1.35 | 0.91 | 1.14 |
| 1H-tetrazole | 5.17 | 5.35 | 3.77 | 5.47 | 3.76 | 3.84 |
| 1H-triazole | 3.29 | 4.22 | 3.40 | 5.23 | 4.19 | 3.93 |
| thioformaldehyde | 4.57 | 3.16 | 2.76 | 4.30 | 1.91 | 2.20 |
| aminoperoxide | 2.46 | 3.07 | 2.49 | 4.93 | 4.18 | 3.97 |
| bipyrimidine | 8.81 | 7.49 | 9.01 | 5.38 | 5.38 | 4.11 |
| hydroxymethylphosphine | 4.79 | 3.15 | 2.97 | 5.06 | 5.80 | 5.97 |
| bis(phosphanyl)hydrazine | 4.29 | 5.09 | 5.41 | 3.29 | 3.84 | 3.99 |

**Table S15**. Average negative log-likelihood values of the model predictions of the spin-splitting energies on all four data sets.

| Model | Training set | Validation set | Composition test set | Ligand test set |
|---|---|---|---|---|
| KRR standard-RACs | 2.84 | 3.84 | 3.80 | 4.55 |
| KRR two-body | 3.77 | 4.34 | 4.40 | 5.82 |
| KRR three-body | 3.14 | 4.41 | 4.60 | 5.97 |
| NN standard-RACs | 17.83 | 29.60 | 27.64 | 16.10 |
| NN two-body | 24.85 | 31.37 | 22.82 | 15.36 |
| NN three-body | 8.16 | 9.63 | 7.97 | 5.18 |

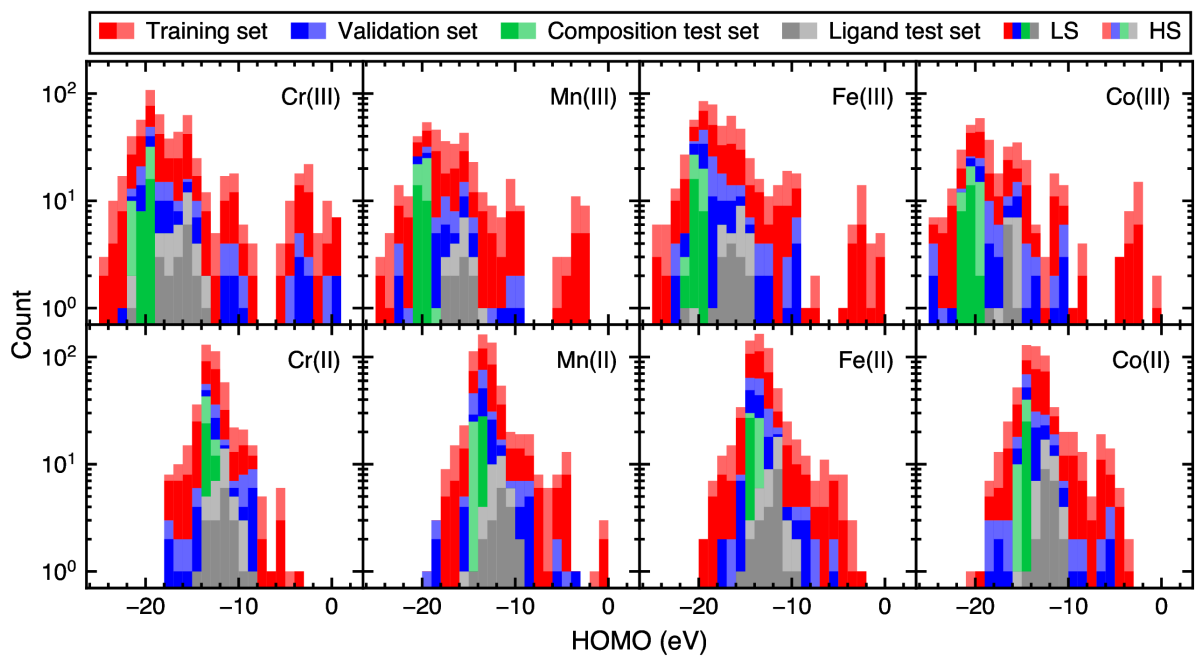

**Figure S8.** Stacked histogram of the DFT calculated HOMO energies for all four data sets and eight metal/oxidation state combinations with a bin width of 1 eV grouped by set origin as indicated in inset legend. Low-spin states are shown in saturated colors, and high-spin states are shown in translucent colors.

**Table S16**. Distribution of total ligand charge for the complexes in the training set for each of the eight metal/oxidation state combinations.

| Total ligand charge | Cr(III) | Cr(II) | Mn(III) | Mn(II) | Fe(III) | Fe(II) | Co(III) | Co(II) |
|---|---|---|---|---|---|---|---|---|
| 0 | 144 | 131 | 84 | 170 | 124 | 180 | 87 | 156 |
| -1 | 22 | 13 | 21 | 21 | 26 | 19 | 13 | 24 |
| -2 | 21 | 4 | 18 | 12 | 27 | 16 | 17 | 19 |
| -3 | 2 | 0 | 3 | 2 | 0 | 0 | 0 | 0 |
| -4 | 26 | 0 | 10 | 0 | 15 | 0 | 14 | 0 |
| -5 | 3 | 0 | 0 | 0 | 0 | 0 | 0 | 0 |

**Table S17**. Coefficients of determination $R^2$ of the model predictions of the combined LS and HS HOMO energies on all four data sets.

| Model | Training set | Validation set | Composition test set | Ligand test set |
|---|---|---|---|---|
| KRR standard-RACs | **0.998** | 0.984 | 0.943 | 0.698 |
| KRR two-body | 0.984 | 0.974 | 0.986 | 0.569 |
| KRR three-body | 0.995 | 0.978 | **0.990** | 0.670 |
| NN standard-RACs | 0.993 | **0.987** | 0.952 | 0.756 |
| NN two-body | 0.984 | 0.975 | 0.986 | 0.696 |
| NN three-body | 0.992 | 0.981 | 0.904 | **0.769** |

**Table S18**. Composition dependence of the ML model mean absolute errors for the combined LS and HS HOMO energies on the composition test set in eV.

| Model | $M(L_A)_6$ | $M(L_A)_5(L_B)_1$ | $M(L_A)_4(L_B)_2$ | $M(L_A)_3(L_B)_3$ |
|---|---|---|---|---|
| KRR standard-RACs | 0.15 | 0.56 | 0.62 | 0.50 |
| KRR two-body | 0.25 | 0.27 | 0.29 | 0.28 |
| KRR three-body | 0.08 | 0.20 | 0.24 | 0.23 |
| NN standard-RACs | 0.39 | 0.53 | 0.58 | 0.56 |
| NN two-body | 0.38 | 0.32 | 0.30 | 0.29 |
| NN three-body | 0.33 | 0.56 | 0.83 | 0.93 |

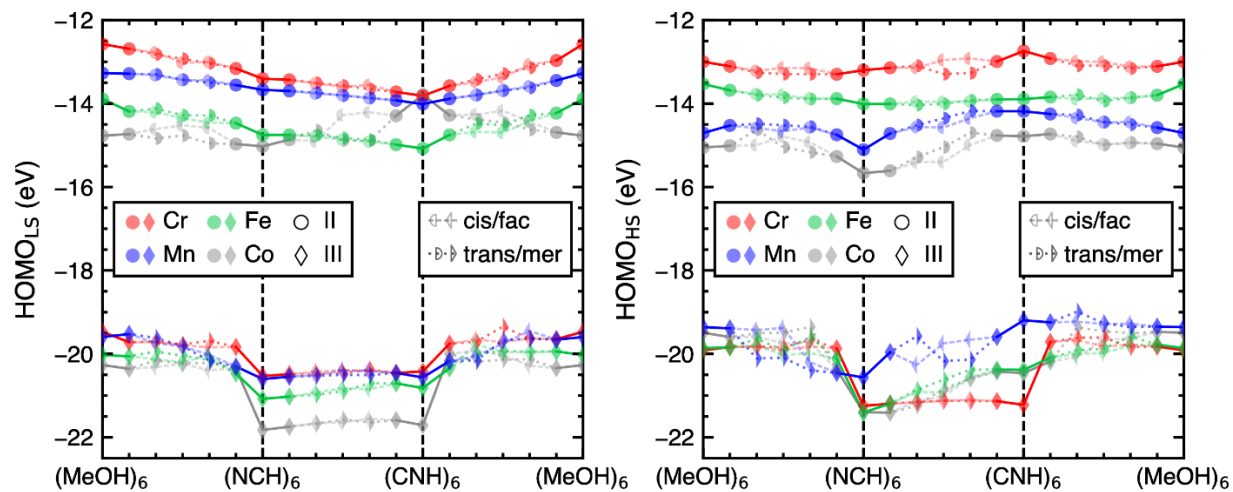

**Figure S9.** Plot of the DFT HOMO energy interpolation curves for binary complexes of three ligands, methanol (MeOH), hydrogen cyanide (NCH), and hydrogen isocyanide (CNH). Neighboring DFT data points corresponding to the same central metal ion are connected by straight lines to guide the eye. For compositions with multiple structural isomers, values are plotted using half markers and dashed/dotted lines.

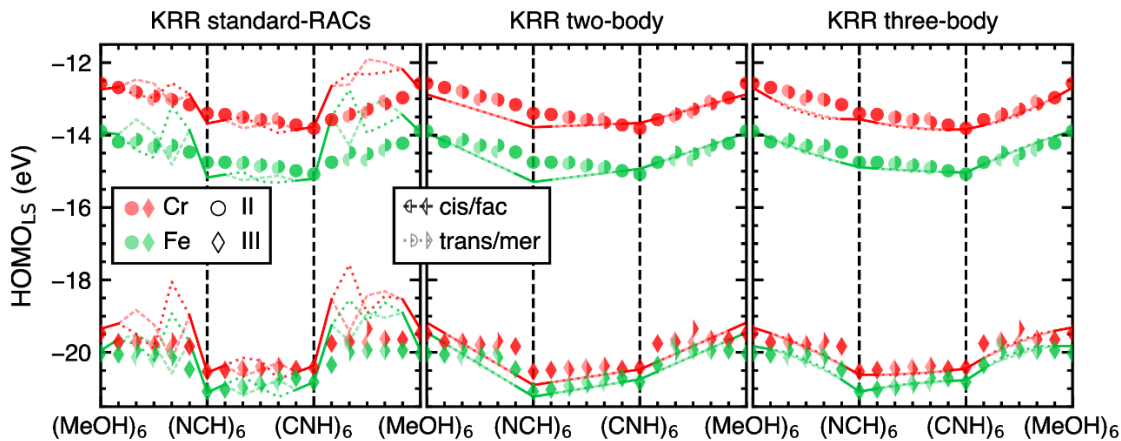

**Figure S10.** Plot of the LS HOMO energy (in eV) interpolation curves for binary complexes of three ligands, methanol (MeOH), hydrogen cyanide (NCH), and hydrogen isocyanide (CNH), and two metals Cr and Fe showing the DFT reference values (scatter) and the corresponding ML predictions (lines) for KRR models using different featurization approaches (columns). For compositions with multiple structural isomers, values are plotted using half markers for the DFT reference and dashed/dotted lines for the ML predictions.

**Table S19.** Mean absolute error of the model predictions of the combined LS and HS HOMO energies on subsets of the ligand test set for each of the eight metal/oxidation state combinations and averaged over all M(II) and M(III) complexes in eV.

| Model | Cr(III) | Mn(III) | Fe(III) | Co(III) | Cr(II) | Mn(II) | Fe(II) | Co(II) |
|---|---|---|---|---|---|---|---|---|
| KRR standard-RACs | 1.27 | 1.35 | 1.22 | 1.67 | 0.76 | 0.81 | 0.84 | 0.95 |
| | 1.34 | | | | 0.85 | | | |
| KRR two-body | 1.66 | 1.43 | 1.52 | 1.83 | 0.96 | 1.03 | 1.07 | 1.16 |
| | 1.60 | | | | 1.06 | | | |
| KRR three-body | 1.39 | 1.36 | 1.28 | 1.65 | 0.80 | 0.93 | 0.92 | 1.10 |
| | 1.39 | | | | 0.94 | | | |
| NN standard-RACs | 1.20 | 1.42 | 1.07 | 1.37 | 0.70 | 0.74 | 0.81 | 0.88 |
| | 1.23 | | | | 0.79 | | | |
| NN two-body | 1.49 | 1.24 | 1.20 | 1.48 | 0.82 | 0.81 | 0.85 | 0.93 |
| | 1.36 | | | | 0.86 | | | |
| NN three-body | 1.17 | 1.23 | 0.97 | 1.13 | 0.78 | 0.84 | 0.88 | 0.99 |
| | 1.12 | | | | 0.88 | | | |

**Table S20.** Mean absolute error of the model predictions for the combined LS and HS HOMO energies on subsets of the ligand test set for each of the 21 ligands in eV. The largest error for each model is indicated in bold.

| Ligand | KRR | | | NN | | |
| --- | --- | --- | --- | --- | --- | --- |
| | standard-RACs | two-body | three-body | standard-RACs | two-body | three-body |
| 4H-pyran | 0.32 | 0.69 | 0.39 | 1.12 | 0.56 | 1.49 |
| ethenediol | 1.63 | 2.22 | 2.08 | 1.40 | 1.95 | 1.16 |
| bifuran | 0.33 | 1.57 | 1.38 | 0.95 | 0.64 | 1.15 |
| pyridine-N-oxide | 0.88 | 1.65 | 1.81 | 1.84 | 1.55 | **2.55** |
| acrylamide | 0.34 | 0.70 | 0.83 | 0.26 | 0.15 | 0.73 |
| dimethylformamide | **2.45** | 0.43 | 0.09 | 1.56 | 0.82 | 0.80 |
| thiophene | 0.73 | 0.45 | 0.13 | 0.12 | 0.39 | 0.16 |
| thiane | 1.59 | 1.26 | 1.10 | 0.76 | 0.90 | 0.47 |
| 4H-thiopyran | 0.37 | 0.56 | 0.92 | 1.65 | 1.14 | 1.56 |
| oxazoline | 2.37 | **2.98** | **2.86** | 1.87 | **2.77** | 2.06 |
| thioazole | 1.17 | 1.64 | 1.80 | 1.34 | 1.72 | 0.82 |
| NHCHOH | 0.31 | 1.69 | 1.35 | 0.63 | 1.55 | 1.01 |
| PHCHOH | 1.08 | 2.88 | 1.99 | 0.34 | 2.24 | 1.20 |
| tetrazane | 0.75 | 0.39 | 0.66 | 0.36 | 0.94 | 0.42 |
| 1H-tetrazole | 2.34 | 1.17 | 1.18 | **2.03** | 1.12 | 0.99 |
| 1H-triazole | 0.89 | 0.64 | 0.53 | 0.60 | 0.51 | 0.94 |
| thioformaldehyde | 0.46 | 0.81 | 0.71 | 0.54 | 0.46 | 0.44 |
| aminoperoxide | 1.09 | 0.26 | 0.28 | 0.61 | 0.25 | 1.02 |
| bipyrimidine | 0.37 | 1.14 | 0.98 | 0.50 | 0.25 | 0.42 |
| hydroxymethylphosphine | 0.54 | 2.68 | 1.57 | 0.52 | 1.24 | 0.42 |
| bis(phosphanyl)hydrazine | 0.51 | 0.50 | 0.50 | 0.65 | 0.71 | 0.69 |

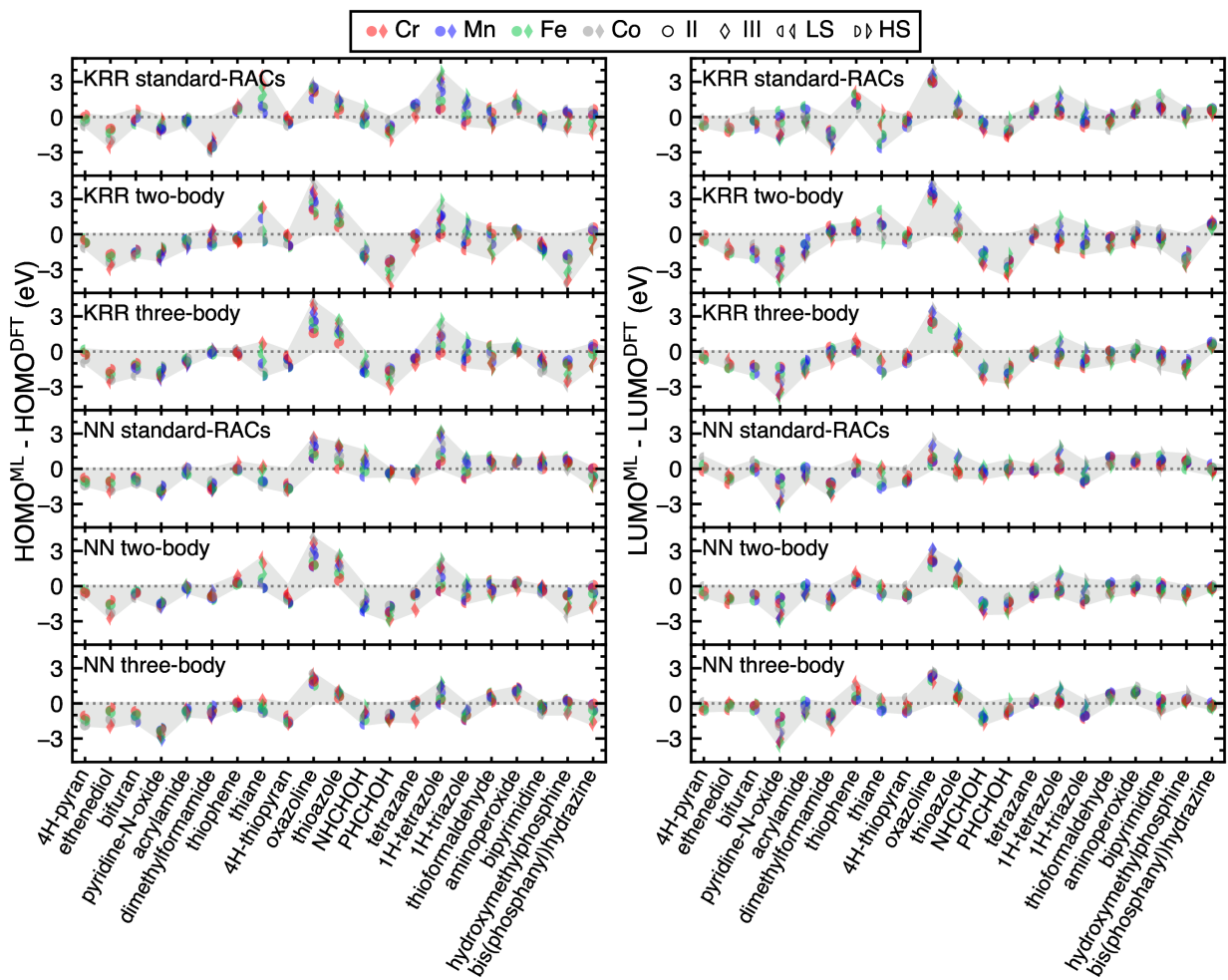

**Figure S11.** Prediction error of the LS and HS HOMO (left) and LUMO (right) energies for all six models on subsets of the ligand test set for each of the 21 ligands in eV. The shaded area indicates the maximum error for each ligand.

**Table S21.** Standard deviation of the absolute model prediction errors for the combined LS and HS HOMO energies on subsets of the ligand test set for each of the 21 ligands in eV. The largest value for each model is indicated in bold.

| | KRR | | | NN | | |
|---|---|---|---|---|---|---|
| Ligand | standard-RACs | two-body | three-body | standard-RACs | two-body | three-body |
| 4H-pyran | 0.32 | 0.69 | 0.39 | 1.12 | 0.56 | 1.49 |
| ethenediol | 1.63 | 2.22 | 2.08 | 1.40 | 1.95 | 1.16 |
| bifuran | 0.33 | 1.57 | 1.38 | 0.95 | 0.64 | 1.15 |
| pyridine-N-oxide | 0.88 | 1.65 | 1.81 | 1.84 | 1.55 | **2.55** |
| acrylamide | 0.34 | 0.70 | 0.83 | 0.26 | 0.15 | 0.73 |
| dimethylformamide | **2.45** | 0.43 | 0.09 | 1.56 | 0.82 | 0.80 |
| thiophene | 0.73 | 0.45 | 0.13 | 0.12 | 0.39 | 0.16 |
| thiane | 1.59 | 1.26 | 1.10 | 0.76 | 0.90 | 0.47 |
| 4H-thiopyran | 0.37 | 0.56 | 0.92 | 1.65 | 1.14 | 1.56 |
| oxazoline | 2.37 | **2.98** | 2.86 | 1.87 | **2.77** | 2.06 |
| thioazole | 1.17 | 1.64 | 1.80 | 1.34 | 1.72 | 0.82 |
| NHCHOH | 0.31 | 1.69 | 1.35 | 0.63 | 1.55 | 1.01 |
| PHCHOH | 1.08 | 2.88 | 1.99 | 0.34 | 2.24 | 1.20 |
| tetrazane | 0.75 | 0.39 | 0.66 | 0.36 | 0.94 | 0.42 |
| 1H-tetrazole | 2.34 | 1.17 | 1.18 | **2.03** | 1.12 | 0.99 |
| 1H-triazole | 0.89 | 0.64 | 0.53 | 0.60 | 0.51 | 0.94 |
| thioformaldehyde | 0.46 | 0.81 | 0.71 | 0.54 | 0.46 | 0.44 |
| aminoperoxide | 1.09 | 0.26 | 0.28 | 0.61 | 0.25 | 1.02 |
| bipyrimidine | 0.37 | 1.14 | 0.98 | 0.50 | 0.25 | 0.42 |
| hydroxymethylphosphine | 0.54 | 2.68 | 1.57 | 0.52 | 1.24 | 0.42 |
| bis(phosphanyl)hydrazine | 0.51 | 0.50 | 0.50 | 0.65 | 0.71 | 0.69 |

**Text S2.** Analysis of the ML predictions for LUMO energies

We investigated the extent to which trends in model performance for HOMO levels also applied to the LUMO set. While all models trained to predict LUMO energies achieve slightly lower MAEs compared to the HOMO results (on average over all models and data sets 0.08 eV lower), the overall trends between different ML methods and featurization approaches remain similar on all four data sets (Supplementary Material Table S22). This observation is at first surprising because energies of unoccupied orbitals might be expected to be noisier than occupied orbitals, as the orbital is only subject to orthogonality during the optimization. The most significant difference for the LUMO results is that the three-body NN does not have a high MAE on the composition test set. We attribute this to overall smoother DFT target data for the LUMO energies (Supplementary Material Figure S14). As a consequence, the average curvature of the three-body NN predictions of 0.12 eV is significantly lower than the value of 0.55 eV obtained for the HOMO energies and the prediction error is almost constant along the entire interpolation curve between homoleptic complexes (Supplementary Material Table S23). An analysis of the 21 ligand subsets for the ligand test set reveals similar trends to the HOMO results where all models exhibit the highest MAEs on M(oxazoline)$_6$ and M(pyridine-N-oxide)$_6$ complexes (Supplementary Material Table S21).

**Table S22**. Mean absolute errors of the model predictions of the combined LS and HS LUMO energies on all four data sets in eV. The smallest error for each set is indicated in bold.

| Model | Training set | Validation set | Composition test set | Ligand test set |
|---|---|---|---|---|
| KRR standard-RACs | **0.11** | 0.48 | 0.47 | 0.94 |
| KRR two-body | 0.25 | 0.36 | 0.31 | 1.19 |
| KRR three-body | 0.12 | 0.32 | **0.22** | 1.01 |
| NN standard-RACs | 0.29 | 0.42 | 0.49 | **0.71** |
| NN two-body | 0.27 | 0.32 | 0.40 | 0.88 |
| NN three-body | 0.25 | **0.31** | 0.44 | 0.80 |

**Table S23**. Composition dependence of the ML model mean absolute errors for the combined LS and HS LUMO energies on the composition test set in eV. The smallest error for each subset is indicated in bold.

| Model | $M(L_A)_6$ | $M(L_A)_5(L_B)_1$ | $M(L_A)_4(L_B)_2$ | $M(L_A)_3(L_B)_3$ |
|---|---|---|---|---|
| KRR standard-RACs | 0.25 | 0.32 | 0.50 | 0.54 |
| KRR two-body | 0.37 | 0.31 | 0.31 | 0.30 |
| KRR three-body | **0.11** | **0.18** | **0.24** | **0.23** |
| NN standard-RACs | 0.46 | 0.44 | 0.49 | 0.53 |
| NN two-body | 0.46 | 0.41 | 0.41 | 0.38 |
| NN three-body | 0.49 | 0.44 | 0.45 | 0.43 |

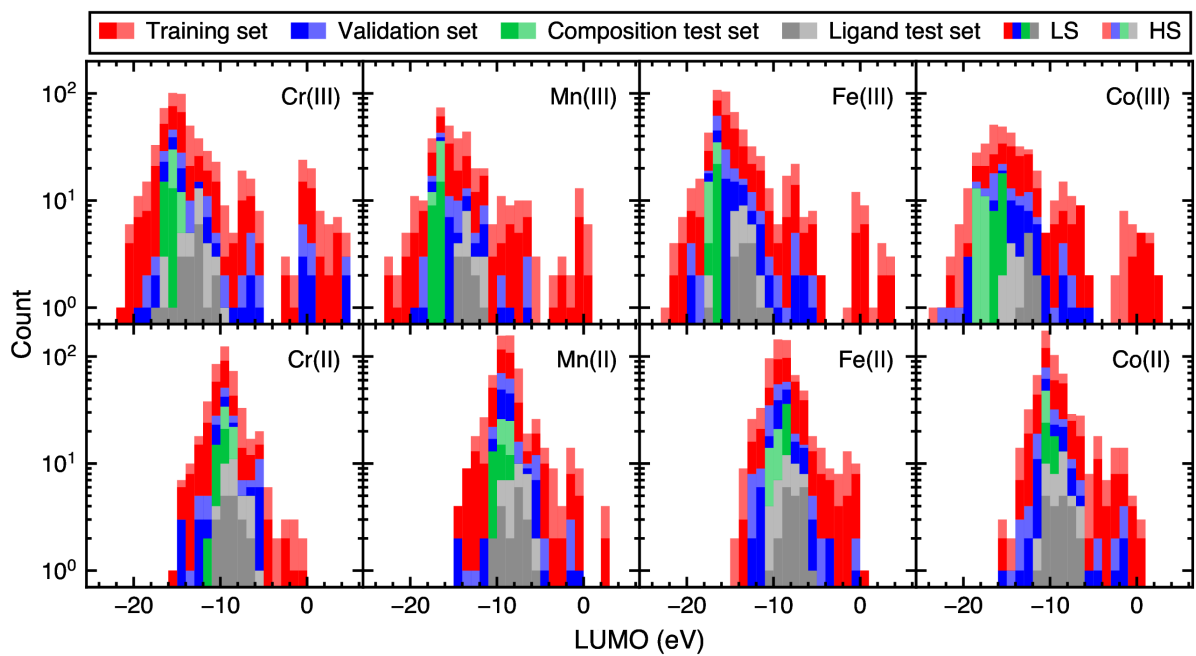

**Figure S12.** Stacked histogram of the DFT calculated LUMO energies for all four data sets and eight metal/oxidation state combinations with a bin width of 1 eV grouped by set origin as indicated in inset legend. Low-spin states are shown in saturated colors, and high-spin states are shown in translucent colors.

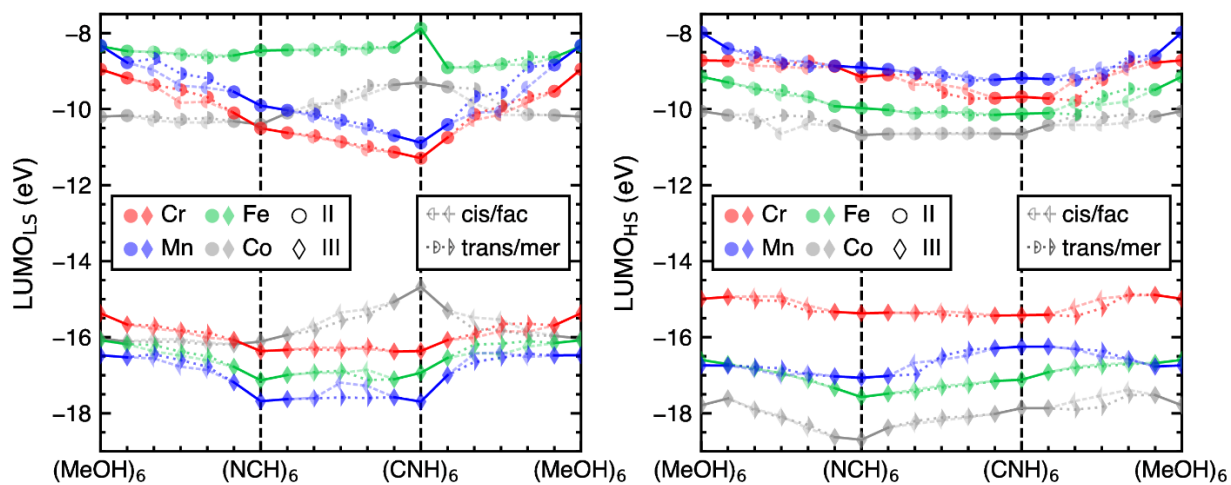

**Figure S13.** Plot of the DFT LUMO energy interpolation curves for binary complexes of three ligands, methanol (MeOH), hydrogen cyanide (NCH), and hydrogen isocyanide (CNH). For compositions with multiple structural isomers values are plotted using half markers and dashed/dotted lines.

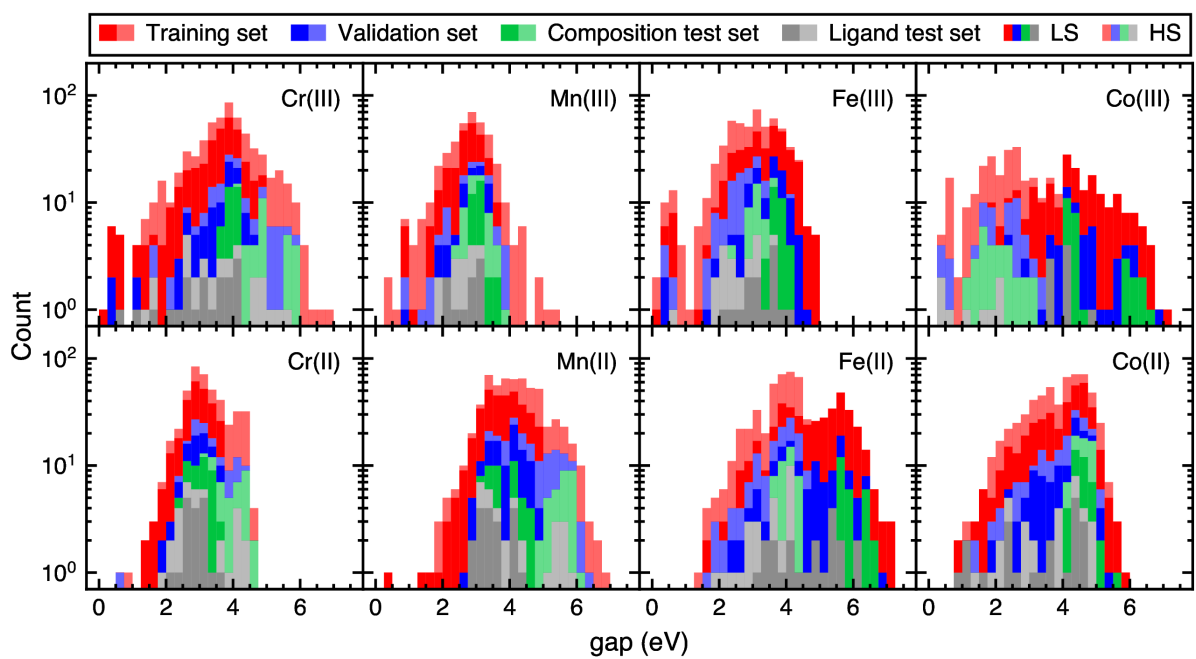

**Figure S14.** Stacked histogram of the DFT calculated HOMO-LUMO gap energies for all four data sets and eight metal/oxidation state combinations with a bin width of 1 eV grouped by set origin as indicated in inset legend. Low-spin states are shown in saturated colors, and high-spin states are shown in translucent colors.

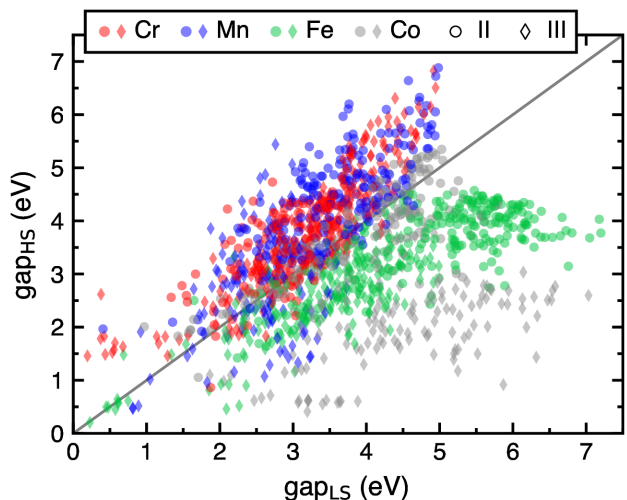

**Figure S15.** Parity plot of the DFT calculated LS HOMO–LUMO gap energies of the training set versus the corresponding HS HOMO–LUMO gap energies grouped by metal and oxidation state as indicated in top legend. For open-shell configurations an energy-based convention for HOMO and LUMO are used, i.e., HOMO=max(HOMO$_\alpha$, HOMO$_\beta$) and LUMO=min(LUMO$_\alpha$, LUMO$_\beta$).

**Table S24.** Pearson correlation coefficient $r$ and coefficients of determination $R^2$ of the LS DFT HOMO–LUMO gap energies versus the HS DFT HOMO–LUMO gap energies on subsets of the training set for each of the eight metal/oxidation state combinations.

| Metric | Cr(III) | Cr(II) | Mn(III) | Mn(II) | Fe(III) | Fe(II) | Co(III) | Co(II) | all |
|---|---|---|---|---|---|---|---|---|---|
| Pearson $r$ | 0.89 | 0.56 | 0.34 | 0.74 | 0.78 | 0.69 | 0.50 | 0.83 | 0.41 |
| $R^2$ score | 0.23 | -1.87 | -1.86 | -0.99 | -0.09 | -1.53 | -5.16 | 0.65 | -0.21 |

**Table S25.** Number of negative ML predictions of the HOMO–LUMO gap energy on all four data sets from a possible 1444 predictions on the training set, 362 predictions on the validation set, 192 predictions on the composition test set, and 127 predictions on the ligand test set.

| Model | Training set | Validation set | Composition test set | Ligand test set |
|---|---|---|---|---|
| KRR standard-RACs | 0 | 0 | 0 | 3 |
| KRR two-body | 0 | 0 | 0 | 0 |
| KRR three-body | 0 | 0 | 0 | 0 |
| NN standard-RACs | 0 | 0 | 0 | 0 |
| NN two-body | 2 | 0 | 0 | 1 |
| NN three-body | 1 | 0 | 0 | 0 |

**Table S26**. Coefficients of determination $R^2$ of the model predictions of the combined LS and HS HOMO–LUMO gap energies on all four data sets. The highest value for each set is indicated in bold.

| Model | Training set | Validation set | Composition test set | Ligand test set |
|---|---|---|---|---|
| KRR standard-RACs | **0.960** | 0.550 | 0.657 | 0.079 |
| KRR two-body | 0.808 | 0.681 | 0.808 | 0.338 |
| KRR three-body | 0.925 | 0.720 | **0.874** | 0.504 |
| NN standard-RACs | 0.848 | 0.731 | 0.639 | 0.454 |
| NN two-body | 0.782 | 0.697 | 0.731 | 0.573 |
| NN three-body | 0.845 | **0.757** | -0.050 | **0.600** |

**Table S27.** Mean absolute errors of the models trained directly on the HOMO–LUMO gap energies for the combined LS and HS HOMO–LUMO gap energies on all four data sets in eV. The lowest error for each set is indicated in bold.

| Model | Training set | Validation set | Composition test set | Ligand test set |
|---|---|---|---|---|
| KRR standard-RACs | **0.13** | 0.41 | 0.49 | **0.57** |
| KRR two-body | 0.31 | 0.44 | **0.25** | 0.63 |
| KRR three-body | 0.13 | **0.38** | 0.62 | 0.62 |
| NN standard-RACs | 0.34 | 0.43 | 0.43 | 0.69 |
| NN two-body | 0.40 | 0.43 | 0.47 | 0.63 |
| NN three-body | 0.32 | 0.38 | 0.96 | 0.62 |